\documentclass[10pt]{article}
\usepackage{multirow}
\usepackage[nottoc]{tocbibind}
\usepackage{geometry}
\usepackage[affil-it, auth-sc]{authblk}
\usepackage[utf8]{inputenc}
\usepackage{graphicx}
\usepackage{amsthm}
\usepackage{amssymb}
\usepackage{amsmath}
\usepackage{color, soul}
\usepackage{verbatim}
\usepackage{lineno}
\usepackage{todonotes}
\usepackage{subcaption}
\usepackage{ulem}
\usepackage{float}
\definecolor{fnalblue}{RGB}{ 0, 76, 151}
\definecolor{fnalbluedark}{RGB}{  0, 40, 85}

\usepackage[pdftex,colorlinks,bookmarks,
linkcolor=fnalbluedark, 
citecolor={blue!50!black},
urlcolor={blue!50!black}]{hyperref}
\usepackage[ampersand]{easylist}
\usepackage{babel}
\usepackage{bm}
\usepackage[sort&compress,numbers]{natbib}
\usepackage{doi}
\usepackage{eso-pic}
\usepackage[capitalise]{cleveref}

\usepackage{physics,slashed}
\usepackage[force]{feynmp-auto}

\presetkeys{todonotes}{inline, color=blue!30, size=\small}{}
\definecolor{darkred}{rgb}{0.9, 0.0, 0.0}
\definecolor{darkgreen}{rgb}{0.0, 0.5, 0.0}
\allowdisplaybreaks

\newcommand{\nl}{\nonumber \\ }
\newcommand{\Order}{{\cal O}}

\newcommand{\e}{{\rm e}}
\newcommand{\iu}{{\rm i}}

\newcommand{\D}{D}

\newcommand{\gammatrad}{\eta}
\newcommand{\vel}{\beta}
\newcommand{\betaFunc}{B}

\makeatletter
\newcommand{\vast}{\bBigg@{4}}
\newcommand{\Vast}{\bBigg@{5}}
\makeatother

\def\arrowang{15}
\def\arrowlen{2.5mm}
\def\decorsize{2.5mm}

\begin{document}

\AddToShipoutPictureFG*{\AtPageUpperLeft{\put(-60,-75){\makebox[\paperwidth][r]{FERMILAB-PUB-23-454-T}}}}
\AddToShipoutPictureFG*{\AtPageUpperLeft{\put(-60,-60){\makebox[\paperwidth][r]{CALT-TH-2023-034}}}}

\title{\Large\bf All orders factorization and the Coulomb problem}
\author[1,2]{Richard~J.~Hill}
\author[1,2,3]{Ryan~Plestid}
\affil[1]{University of Kentucky, Department of Physics and Astronomy, Lexington, KY 40506 USA \vspace{1.2mm}}
\affil[2]{Fermilab, Theoretical Physics Department, Batavia, IL 60510, USA
\vspace{1.2mm}}
\affil[3]{Walter Burke Institute for Theoretical Physics, California Institute of Technology, Pasadena, CA, 91125 USA\vspace{1.2mm}}

\date{\today}

\maketitle
\begin{abstract}

\noindent 
In the limit of large nuclear charge, $Z\gg 1$, or small lepton velocity, $\vel \ll 1$, Coulomb corrections to nuclear beta decay and related processes are enhanced as $Z\alpha/\vel$ and become large or even non-perturbative (with $\alpha$ the QED fine structure constant).    We provide a constructive demonstration of factorization to all orders in perturbation theory for these processes and compute the all-orders hard and soft functions appearing in the factorization formula. We clarify the relationship between effective field theory amplitudes and historical treatments of beta decay in terms of a Fermi function.
\end{abstract}

\newpage
\tableofcontents
\newpage


\section{Introduction}
The Coulomb field of a nucleus can have dramatic consequences for low energy phenomena. Relative to other QED effects, Coulomb corrections are large because {\it i)} they are enhanced by the charge of the nucleus \cite{Sommerfeld:1931qaf,Fermi:1934hr,Davies:1954zz,Bethe:1954zz}, {\it ii)} they are enhanced at low relative velocity \cite{Sakharov:1948plh,Fadin:1990wx,Kulesza:2008jb,Beneke:2010da}, and {\it iii)} loop integrals receive systematic $\pi$-enhancements \cite{largepi}. Many precision  experiments involve leptons interacting with nuclei 
\cite{Nunokawa:2007qh,Hyper-KamiokandeWorkingGroup:2014czz,Diwan:2016gmz,NOvA:2021nfi,T2K:2021xwb,DUNE:2020jqi,deGouvea:2013zba,Bernstein:2013hba,Lee:2018wcx,Bernstein:2019fyh,Bopp:1986rt,Ando:2004rk,Darius:2017arh,Seng:2018yzq,Seng:2018qru,Fry:2018kvq,Czarnecki:2019mwq,Hayen:2020cxh,Seng:2020wjq,Gorchtein:2021fce,UCNt:2021pcg,Shiells:2020fqp,Cirigliano:2022hob,Cirigliano:2013xha,Glick-Magid:2016rsv,Gonzalez-Alonso:2018omy,Hardy:2020qwl,Glick-Magid:2021uwb,Falkowski:2021vdg,Brodeur:2023eul,Crivellin:2020ebi,Coutinho:2019aiy,Crivellin:2021njn,Crivellin:2020lzu,Cirigliano:2022yyo}, and require the systematic treatment of Coulomb corrections and their interplay with other subleading effects \cite{Jaus:1970tah,Jaus:1972hua,Sirlin:1986cc,Jaus:1986te,Cirigliano:2023fnz,Hill:2023acw,z2a3anom}.

Factorization theorems underlie much of our ability to retain theoretical control in precision measurements involving nucleons, nuclei, and other hadrons \cite{Collins:1989gx,Bodwin:1994jh,Bauer:2002nz,Hill:2010yb,Hill:2014yxa,Hill:2016gdf}. Factorization arises from the separation of different energy scales 
involved in a physical process, with the components in the factorization formula identified with contributions from each scale \cite{Beneke:1997zp,Jantzen:2011nz}.  In terms of a sequence of effective field theories (EFTs), the components are identified as the corresponding sequence of matching coefficients, and the final low-energy matrix element.  Historically, Coulomb corrections have been understood not in terms of EFT, but by appealing to wavefunction methods i.e., solutions of the Dirac or Schrodinger equation \cite{Fermi:1934hr,Wilkinson:1982hu,Kotila:2012zza,Hayen:2017pwg}. 
Such wavefunction descriptions contain the correct long-distance behavior, which is however, 
intertwined with model-dependent short-distance behavior. Separating scales allows us to 
systematically resum logarithms and study higher order radiative corrections using standard tools of effective field theory. 
The interplay of high order Coulomb corrections with other subleading effects is crucial for precision measurements, and in particular for nuclear beta decays~\cite{Hill:2023acw}.

In this paper, we demonstrate factorization for radiative corrections induced by photon exchange between charged leptons and a static Coulomb field, and compute explicit all-orders expressions for the components of the factorization formula. We describe how traditional wavefunction methods are related to dimensionally regulated Feynman integrals order by order in perturbation theory. Using this correspondence, and a new all-orders calculation of the short-distance region, we extract the universal $\overline{\rm MS}$ Coulomb corrections to the matrix element for a contact interaction (as is relevant for nuclear beta decays) to all orders in perturbation theory. 

The remainder of the paper is organized as follows. \cref{sec:coulomb-problem} introduces notation for Coulomb corrections from a diagrammatic perspective. \cref{sec:schro} considers the Schrodinger-Coulomb problem and establishes the correspondence between wavefunctions and the diagrammatic expansions. \Cref{sec:dirac} considers the Dirac-Coulomb problem and extracts the relevant EFT matrix element to all orders in perturbation theory. \Cref{sec:discuss} highlights new and interesting features of the preceding analysis and comments on phenomenological applications. 

\section{Coulomb corrections and contact interactions \label{sec:coulomb-problem} }
Consider a reaction that takes place via an effective contact interaction in the vicinity of a heavy particle with charge $Z$. The outgoing charged particles (``leptons") can exchange photons with the heavy particle (``nucleus").  For neutral current processes (i.e., when the initial and final nuclear states have the same charge $Z$), QED radiative corrections can be straightforwardly organized as 
a series in powers of 
$\alpha$, $Z\alpha$ and $Z^2\alpha$, 
with each power being separately QED gauge invariant.\!\footnote{For charged current process, e.g.,
$A[Z+1] \to B[Z] + \ell^+ + \nu_\ell$,
the same Coulomb factor describes 
the leading $Z$-enhanced contributions. See Ref.~\cite{eikonal_algebra} for a discussion of how subleading contributions are organized.  
} We will consider the static limit, in which the particle of charge $Z$ in the initial and final state is heavy, so that recoil corrections can be neglected. For low momentum probes satisfying $|\vb{p}| \ll 1/R$ with $R$ the charge radius of the heavy particle, the point-like limit is applicable and universal corrections to the amplitude can be computed using Feynman rules for a static external Coulomb field \cite{Weinberg:1995mt}. In this static limit, terms $\sim Z^m \alpha^n$ vanish for $m>n$~\cite{Beg:1969zu}.  In the following we consider the leading series of terms, $\sim (Z\alpha)^n$, for $n\ge 0$.

As an explicit example, consider di-lepton production via a short-range neutral current in some nuclear decay:\footnote{An electromagnetic $E0$ transition can mimic the same phenomenology if both $e^+$ and $e^-$ are non-relativistic, such that the virtual photon that mediates the transition is far off-shell.}
\begin{equation}\label{eq:AB}
    A(v_A) \rightarrow B(v_B)~+ \ell^-(\vb{p}_1) ~+ \ell^+(\vb{p}_2)~,
\end{equation}
where states $A$ and $B$ have charge $Z$, and $v_B^\mu=v_A^\mu=v^\mu = (1,\vb{0})$
which defines the static limit. As discussed above, this can be reduced to an external
field problem describing the production of a di-lepton pair in a static Coulomb field,
\begin{equation}
    \label{equality-1}
    \begin{split}
    \raisebox{30pt}{$\mathcal{M}=$}& \!\!\!\!\!\!
    \begin{fmffile}{term-1}
    \fmfset{arrow_ang}{\arrowang}
    \fmfset{arrow_len}{\arrowlen}
    \fmfset{decor_size}{\decorsize}
        \begin{fmfgraph*}(65,65) 
            \fmfleft{i1}
            \fmfbottom{b0,b1,b2,b3}
            \fmfright{o0,o1,o2,o3}
            \fmf{phantom,tension=3}{i1,v1}
            \fmfv{d.sh=square}{v1}
            \fmf{fermion}{o1,v1,o2}
        \end{fmfgraph*}
    \end{fmffile}
    \raisebox{30pt}{$~~+~~\Vast($}
    \hspace{-12pt}
    \begin{fmffile}{term-2a}
    \fmfset{arrow_ang}{\arrowang}
    \fmfset{arrow_len}{\arrowlen}
    \fmfset{decor_size}{\decorsize}
        \begin{fmfgraph*}(65,65) 
            \fmfleft{i1}
            \fmfbottom{b0,b1,b2,b3}
            \fmfright{o0,o1,o2,o3}
            \fmf{phantom,tension=3}{i1,v1}
            \fmfv{d.sh=square}{v1}
            \fmf{fermion}{o1,v2,v1,o2}
            \fmf{photon,tension=0}{b2,v2}
            \fmfv{d.sh=cross}{b2}    
        \end{fmfgraph*}
    \end{fmffile}
    \raisebox{30pt}{$~~+$}
    \begin{fmffile}{term-2b}
    \fmfset{arrow_ang}{\arrowang}
    \fmfset{arrow_len}{\arrowlen}
    \fmfset{decor_size}{\decorsize}
        \begin{fmfgraph*}(75,65) 
            \fmfleft{i1}
            \fmftop{b0,b1,b2,b3}
            \fmfright{o0,o1,o2,o3}
            \fmf{phantom,tension=3}{i1,v1}
            \fmfv{d.sh=square}{v1}
            \fmf{fermion}{o1,v1,v2,o2}
            \fmf{photon,tension=0}{b2,v2}
            \fmfv{d.sh=cross}{b2}
        \end{fmfgraph*}
    \end{fmffile}
    \raisebox{30pt}{$\Vast)$}\\
    &\hspace{0.05\linewidth}\raisebox{30pt}{$~~+~~\Vast($}
    \hspace{-12pt}
    \begin{fmffile}{term-3a}
    \fmfset{arrow_ang}{\arrowang}
    \fmfset{arrow_len}{\arrowlen}
    \fmfset{decor_size}{\decorsize}
        \begin{fmfgraph*}(75,65) 
            \fmfleft{i1}
            \fmfbottom{b0,b1,b2,b3,b4}
            \fmfright{o0,o1,o2,o3}
            \fmf{phantom,tension=3.6}{i1,v1}
            \fmfv{d.sh=square}{v1}
            \fmf{fermion}{o1,v3,v2,v1,o2}
            \fmf{photon,tension=0}{b2,v2}
            \fmf{photon,tension=0}{b3,v3}
            \fmfv{d.sh=cross}{b2,b3}
        \end{fmfgraph*}
    \end{fmffile}
    \raisebox{30pt}{$~~+$}
    \begin{fmffile}{term-3b}
    \fmfset{arrow_ang}{\arrowang}
    \fmfset{arrow_len}{\arrowlen}
    \fmfset{decor_size}{\decorsize}
        \begin{fmfgraph*}(75,65) 
            \fmfleft{i1}
            \fmftop{b0,b1,b2,b3,b4}
            \fmfright{o0,o1,o2,o3}
            \fmf{phantom,tension=3.6}{i1,v1}
            \fmfv{d.sh=square}{v1}
            \fmf{fermion}{o1,v1,v2,v3,o2}
            \fmf{photon,tension=0}{b2,v2}
            \fmf{photon,tension=0}{b3,v3}
           \fmfv{d.sh=cross}{b2,b3}
        \end{fmfgraph*}
    \end{fmffile}
    \raisebox{30pt}{$~~+~~$}
    \hspace{-12pt}
    \begin{fmffile}{term-3c}
    \fmfset{arrow_ang}{\arrowang}
    \fmfset{arrow_len}{\arrowlen}
    \fmfset{decor_size}{\decorsize}
        \begin{fmfgraph*}(75,65) 
            \fmfleft{i1}
            \fmfbottom{b0,b1,b2,b3}
            \fmftop{t0,t1,t2,t3}
            \fmfright{o0,o1,o2,o3}
            \fmf{phantom,tension=2}{i1,v1}
            \fmfv{d.sh=square}{v1}
            \fmf{fermion}{o1,v2,v1,v3,o2}
            \fmf{photon,tension=0}{b2,v2}
            \fmf{photon,tension=0}{t2,v3}
           \fmfv{d.sh=cross}{t2,b2}
        \end{fmfgraph*}
    \end{fmffile}
    \raisebox{30pt}{$\Vast)$}
   \raisebox{30pt}{$~+~\dots \,.$}
    \end{split}
\end{equation}
For a contact interaction, the Coulomb corrections on each leg factorize. 
For example, 
\begin{fmffile}{example}
    \fmfset{arrow_ang}{\arrowang}
    \fmfset{arrow_len}{\arrowlen}
    \fmfset{decor_size}{\decorsize}
\begin{equation}
       \parbox{30mm}{
    \begin{fmfgraph*}(75,65) 
            \fmfleft{i1}
            \fmfbottom{b0,b1,b2,b3}
            \fmftop{t0,t1,t2,t3}
            \fmfright{o0,o1,o2,o3}
            \fmf{phantom,tension=2}{i1,v1}
            \fmfv{d.sh=square}{v1}
            \fmf{fermion}{o1,v2,v1,v3,o2}
            \fmf{photon,tension=0}{b2,v2}
            \fmf{photon,tension=0}{t2,v3}
           \fmfv{d.sh=cross}{t2,b2}
        \end{fmfgraph*}
        }
   = 
       \left( \frac{\parbox{25mm}{\hspace{-5mm} \begin{fmfgraph*}(75,65) 
            \fmfleft{i1}
            \fmftop{b0,b1,b2,b3}
            \fmfright{o0,o1,o2,o3}
            \fmf{phantom,tension=3}{i1,v1}
            \fmfv{d.sh=square}{v1}
            \fmf{fermion}{o1,v1,v2,o2}
            \fmf{photon,tension=0}{b2,v2}
            \fmfv{d.sh=cross}{b2}
        \end{fmfgraph*}}
        }{\parbox{25mm}{\hspace{-5mm}
 \begin{fmfgraph*}(75,65) 
            \fmfleft{i1}
            \fmfbottom{b0,b1,b2,b3}
            \fmfright{o0,o1,o2,o3}
            \fmf{phantom,tension=3}{i1,v1}
            \fmfv{d.sh=square}{v1}
            \fmf{fermion}{o1,v1,o2}
        \end{fmfgraph*}}
}\right)
\times
       \parbox{25mm}{
        \hspace{-5mm}
        \begin{fmfgraph*}(75,65) 
            \fmfleft{i1}
            \fmfbottom{b0,b1,b2,b3}
            \fmfright{o0,o1,o2,o3}
            \fmf{phantom,tension=3}{i1,v1}
            \fmfv{d.sh=square}{v1}
            \fmf{fermion}{o1,v1,o2}
        \end{fmfgraph*}
        \hspace{-5mm}
        }
\times
 \left( \frac
        {\parbox{25mm}{\hspace{-5mm} 
         \begin{fmfgraph*}(75,65) 
            \fmfleft{i1}
            \fmfbottom{b0,b1,b2,b3}
            \fmfright{o0,o1,o2,o3}
            \fmf{phantom,tension=3}{i1,v1}
            \fmfv{d.sh=square}{v1}
            \fmf{fermion}{o1,v2,v1,o2}
            \fmf{photon,tension=0}{b2,v2}
            \fmfv{d.sh=cross}{b2}    
        \end{fmfgraph*}\vspace{2mm}}
}
{\parbox{25mm}{\hspace{-5mm}
 \begin{fmfgraph*}(75,65) 
            \fmfleft{i1}
            \fmfbottom{b0,b1,b2,b3}
            \fmfright{o0,o1,o2,o3}
            \fmf{phantom,tension=3}{i1,v1}
            \fmfv{d.sh=square}{v1}
            \fmf{fermion}{o1,v1,o2}
        \end{fmfgraph*}}
        }
\right)
   \,.
\end{equation}
   \end{fmffile}    
In general, the Coulomb corrected matrix element can be represented as 
\begin{fmffile}{factor}
    \fmfset{arrow_ang}{\arrowang}
    \fmfset{arrow_len}{\arrowlen}
    \fmfset{decor_size}{\decorsize}
\begin{equation}
    \label{equality-2}
    \begin{split}
    {\cal M} &= 
       \left( \frac{
        \parbox{25mm}{
        \hspace{-5mm}
        \begin{fmfgraph*}(75,65) 
            \fmfleft{i1}
            \fmfbottom{b0,b1,b2,b3}
            \fmfright{o0,o1,o2,o3}
            \fmf{phantom,tension=3}{i1,v1}
            \fmfv{d.sh=square}{v1}
            \fmf{fermion}{o1,v1,o2}
        \end{fmfgraph*}
        \hspace{-5mm}
        }
        +
       \parbox{25mm}{\hspace{-5mm} \begin{fmfgraph*}(75,65) 
            \fmfleft{i1}
            \fmftop{b0,b1,b2,b3}
            \fmfright{o0,o1,o2,o3}
            \fmf{phantom,tension=3}{i1,v1}
            \fmfv{d.sh=square}{v1}
            \fmf{fermion}{o1,v1,v2,o2}
            \fmf{photon,tension=0}{b2,v2}
            \fmfv{d.sh=cross}{b2}
        \end{fmfgraph*}}
+ \parbox{25mm}{\hspace{-5mm} 
 \begin{fmfgraph*}(75,65) 
            \fmfleft{i1}
            \fmftop{b0,b1,b2,b3,b4}
            \fmfright{o0,o1,o2,o3}
            \fmf{phantom,tension=3.6}{i1,v1}
            \fmfv{d.sh=square}{v1}
            \fmf{fermion}{o1,v1,v2,v3,o2}
            \fmf{photon,tension=0}{b2,v2}
            \fmf{photon,tension=0}{b3,v3}
           \fmfv{d.sh=cross}{b2,b3}
        \end{fmfgraph*}
}+ \dots 
        }{\parbox{25mm}{\hspace{-5mm}
 \begin{fmfgraph*}(75,65) 
            \fmfleft{i1}
            \fmfbottom{b0,b1,b2,b3}
            \fmfright{o0,o1,o2,o3}
            \fmf{phantom,tension=3}{i1,v1}
            \fmfv{d.sh=square}{v1}
            \fmf{fermion}{o1,v1,o2}
        \end{fmfgraph*}}
}\right) \times 
  \parbox{25mm}{
        \hspace{-5mm}
        \begin{fmfgraph*}(75,65) 
            \fmfleft{i1}
            \fmfbottom{b0,b1,b2,b3}
            \fmfright{o0,o1,o2,o3}
            \fmf{phantom,tension=3}{i1,v1}
            \fmfv{d.sh=square}{v1}
            \fmf{fermion}{o1,v1,o2}
        \end{fmfgraph*}
        \hspace{-5mm}
        }
\\
&
\hspace{0.25\linewidth} \times
 \left( \frac
        {
         \parbox{25mm}{
        \hspace{-5mm}
        \begin{fmfgraph*}(75,65) 
            \fmfleft{i1}
            \fmfbottom{b0,b1,b2,b3}
            \fmfright{o0,o1,o2,o3}
            \fmf{phantom,tension=3}{i1,v1}
            \fmfv{d.sh=square}{v1}
            \fmf{fermion}{o1,v1,o2}
        \end{fmfgraph*}
        \hspace{-5mm}
        }
        +
        \parbox{25mm}{\hspace{-5mm} 
         \begin{fmfgraph*}(75,65) 
            \fmfleft{i1}
            \fmfbottom{b0,b1,b2,b3}
            \fmfright{o0,o1,o2,o3}
            \fmf{phantom,tension=3}{i1,v1}
            \fmfv{d.sh=square}{v1}
            \fmf{fermion}{o1,v2,v1,o2}
            \fmf{photon,tension=0}{b2,v2}
            \fmfv{d.sh=cross}{b2}    
        \end{fmfgraph*}\vspace{2mm}}
        +
 \parbox{25mm}{\hspace{-5mm} 
         \begin{fmfgraph*}(75,65) 
            \fmfleft{i1}
            \fmfbottom{b0,b1,b2,b3,b4}
            \fmfright{o0,o1,o2,o3}
            \fmf{phantom,tension=3.6}{i1,v1}
            \fmfv{d.sh=square}{v1}
            \fmf{fermion}{o1,v3,v2,v1,o2}
            \fmf{photon,tension=0}{b2,v2}
            \fmf{photon,tension=0}{b3,v3}
            \fmfv{d.sh=cross}{b2,b3}
        \end{fmfgraph*}
        }
+ \dots 
}
{\parbox{25mm}{\hspace{-5mm}
 \begin{fmfgraph*}(75,65) 
            \fmfleft{i1}
            \fmfbottom{b0,b1,b2,b3}
            \fmfright{o0,o1,o2,o3}
            \fmf{phantom,tension=3}{i1,v1}
            \fmfv{d.sh=square}{v1}
            \fmf{fermion}{o1,v1,o2}
        \end{fmfgraph*}}
        }
\right)   \,.
    \end{split}
\end{equation}
 \end{fmffile}    
We may therefore, without loss of generality, study the Coulomb corrections on a single leptonic leg: any process with multiple leptons that is mediated by a hard current reduces to a product of Coulomb corrections on each leg. 
The perturbative series for a relativistic lepton contains non-trivial Dirac structure that must be inserted between external polarization spinors. 
For the process (\ref{eq:AB}), \cref{equality-2} becomes 
\begin{equation}
{\cal M} = 
\sum_{ijkl} \bar{u}(\vb{p}_1)_i ( \dots )_{ij} \Gamma^{\rm tree}_{jk} (\dots )_{kl} v(\vb{p}_2)_l  \,.
\end{equation}

Traditional analyses of beta decay are 
 expressed in terms of position-space Coulomb wavefunctions 
 for the leptons evaluated at the origin of coordinates, $\psi({\bm{r}}=0)$ \cite{Fermi:1934hr,Wilkinson:1982hu,Kotila:2012zza,Hayen:2017pwg}. As we discuss in \cref{app:wf-schr,app:wf-dirac},
the diagrammatic series represented in \cref{equality-2} is equivalent to a wavefunction solution. However, starting at two-loop order the wavefunction 
$\psi({\bm{r}}=0)$ is UV divergent.\!\footnote{An exception is the non-relativistic Schrodinger Coulomb wavefunction, which is UV finite to all orders.}   
The amplitude must be renormalized and matched consistently with the underlying contact interaction. In order to execute this program, we phrase the problem in terms of factorization of momentum space amplitudes, using dimensional regularization in the $\overline{\rm MS}$ scheme.  Coulomb corrected amplitudes can then be matched consistently to underlying quark-level Lagrangians, and model-dependent position-space wavefunctions are replaced by a systematic expansion in EFT operators.

\section{Schrodinger-Coulomb problem \label{sec:schro}}

Consider the quantum mechanical corrections to a tree-level process for a final-state particle of mass $m$ and electric charge $(-e)$ scattering 
from a Coulomb potential with source charge $(+Ze)$: (we suppress the overall tree-level amplitude factor)
\begin{equation}
    \begin{split}
    \label{eq:SCgeneral}
    {\cal M} &= \sum_{n=0}^\infty {\cal M}^{(n)} 
    = 
    \sum_{n=0}^\infty(2m Z e^2)^n \int {\dd^\D L_1 \over (2\pi)^\D} \int {\dd^\D L_2 \over (2\pi)^\D}\cdots  \int {\dd^\D L_n \over (2\pi)^\D} 
    {1\over \vb{L}_1^2 + \lambda^2}{1\over (\vb{L}_1 -\vb{p})^2 -\vb{p}^2 -\iu 0}
    \\
    &\qquad\qquad\qquad\qquad\qquad \times  {1\over (\vb{L}_1 - \vb{L}_2)^2 + \lambda^2} 
    {1\over (\vb{L}_2 -\vb{p})^2 -\vb{p}^2 -\iu 0} \cdots
     {1\over (\vb{L}_{n-1} - \vb{L}_{n})^2 + \lambda^2} 
    {1\over (\vb{L}_n -\vb{p})^2 -\vb{p}^2 -\iu 0}  \,.
    \end{split}
\end{equation}
Here, $\D=3-2\epsilon$ is the spatial dimension with dimensional regularization parameter $\epsilon$, and $\lambda$ is a photon mass regulator.  The Schrodinger-Coulomb problem describes the 
limit $p \ll \Lambda_{\rm UV} \ll m$, where 
$\Lambda_{\rm UV} ~\sim R^{-1}$ 
denotes the scale of nuclear or hadronic structure.

The amplitude (\ref{eq:SCgeneral}) can be evaluated at each order in perturbation theory. With the photon mass regulator in place, the integrals in \cref{eq:SCgeneral} are UV and IR finite at $\epsilon\to 0$.  
By convention ${\cal M}^{(0)}=1$ and at 
one-loop order,
\begin{align}
     {\cal M}^{(1)}
     &= 
     2mZe^2 \int {\dd^\D L_1 \over (2\pi)^\D}
     {1\over {\vb{L}}^2 + \lambda^2}{1\over (\vb{L} -\vb{p})^2 -\vb{p}^2 -\iu 0 }
     \to {\iu m\over p} {Ze^2\over 4\pi} \left( \log{2p\over \lambda} - {\iu\pi \over 2} \right)  \,,
\end{align}
where the final expression denotes the limit $\epsilon \to 0$, $\lambda\to 0$. 

\subsection{Factorization}

Two momentum regions \cite{Beneke:1997zp,Jantzen:2011nz} are relevant in the integrals (\ref{eq:SCgeneral}): the soft region with $|\vb{L}| \sim \lambda$; and the hard region with $|\vb{L}| \sim p$. Neglecting power corrections in $\lambda/p$, the amplitude may be written 
\begin{align}\label{eq:SCfac}
 {\cal M} &= {\cal M}_S {\cal M}_H \,. 
\end{align} 
In the language of effective operators, ${\cal M}_H$ represents a matching coefficient and ${\cal M}_S$ represents a low-energy operator matrix element, when the full theory represented by \cref{eq:SCgeneral} is matched onto a low-energy theory containing only soft degrees of freedom.\!\footnote{In applications, 
IR divergences are regulated by
physical scales associated with e.g., bremsstrahlung radiation or 
screening effects from 
atomic electrons~\cite{Hayen:2017pwg}. 
It is interesting to note that a photon mass mimics the Yukawa potential typical of the Thomas-Fermi model of atomic screening~\cite{Schiff:1951zza,Tsai:1973py}.
}

\subsection{Soft factor}

The soft limit of \cref{eq:SCgeneral} is readily seen to exponentiate, yielding the soft factor to all orders \cite{Yennie:1961ad,Weinberg:1965nx},
\begin{align}\label{eq:SCsoft}
{\cal M}_S^{(n)} &=  {1\over n! } \left( {\cal M}_S^{(1)} \right)^n
\,,
\end{align}
where the one-loop result is 
\begin{align}\label{eq:SC1loopsoft}
      {\cal M}_S^{(1)} &= 
      2mZe^2\, \int{ \dd^\D L \over (2\pi)^\D}
         {   1 \over \vb{L}^2 + \lambda^2} {1 \over -2\vb{p}\cdot\vb{L} -\iu 0}
         = {\iu m\over p}
         {Ze^2\over (4\pi)^{1-\epsilon} } \Gamma(1+\epsilon) \lambda^{-2\epsilon} {1\over 2\epsilon} \,.
\end{align}
%

\subsection{Hard factor}

The hard factor can similarly be evaluated explicitly, order by order in perturbation theory.  
The hard momentum region is isolated by expanding at 
$\vb{L}^2\gg \lambda^2$. 
At first order, 
\begin{equation}
    \begin{split}
    \label{eq:SC1loophard}
    {{\cal M}_H^{(1)}} &= 2mZe^2 \int {\dd^{\D} L \over (2\pi)^{\D}} {1\over \vb{L}^2}{1\over (\vb{L} -\vb{p})^2 -\vb{p}^2 -\iu 0}
    = {\iu m \over p}{Ze^2\over 4\pi}\left[ {(16\pi)^\epsilon \Gamma(\frac12 +\epsilon)\over \sqrt{\pi}}\right] (-4p^2-\iu 0)^{-\epsilon} \left(-1\over 2\epsilon\right)
    \\
    &= \left[ {\iu Z\overline{\alpha} \over \vel} (- 4p^2/\mu^2 - \iu 0)^{-\epsilon} \right]\left[ -1\over 2\epsilon \right]
    \,,
    \end{split}
\end{equation}
where\footnote{For the relativistic case, we use $\vel=p/E$ to denote the usual relativistic velocity. 
} $\vel = p/m$ and the $\overline{\rm MS}$ coupling $\overline{\alpha}$  is related to the bare charge $e$ in $\D=3-2\epsilon$ dimensions as%
\footnote{ 
Other common definitions in the literature are 
$\mu^{2\epsilon} 4\pi \overline{\alpha}(\mu)/e^2 = (4\pi)^\epsilon\Gamma(1+\epsilon)$ 
or 
$\mu^{2\epsilon} 4\pi \overline{\alpha}(\mu)/e^2 = (4\pi)^\epsilon \exp(-\gamma_{\rm E} \epsilon)$.
The choice in \cref{eq:msbar} is convenient for expressions arising from loop integrals in three dimensions.
These definitions only differ at order $\epsilon^2$
and therefore yield identical expressions for the renormalized amplitudes that we consider. 
}
\begin{align}\label{eq:msbar}
  \mu^{2\epsilon} \overline{\alpha}(\mu) &= 
     {e^2\over 4\pi}  \left[ {(16\pi)^\epsilon \Gamma(\frac12 +\epsilon)\over \sqrt{\pi}}\right] \,. 
\end{align}
At $\epsilon\to 0$, it is readily seen that 
\begin{align}
 {\cal M}^{(1)} = {\cal M}_S^{(1)} + {\cal M}_H^{(1)} \,. 
\end{align}
At second order
\begin{equation}
    \begin{split}
    \label{eq:SC2loophard}
   {{\cal M}_H^{(2)} } &= (2mZe^2)^2 \int {\dd^\D L_1 \over (2\pi)^\D} \int {\dd^\D L_2 \over (2\pi)^\D} 
    {1\over \vb{L}_1^2}{1\over (\vb{L}_1 -\vb{p})^2 -\vb{p}^2 -\iu 0 } {1\over (\vb{L}_1 - \vb{L}_2)^2} 
    {1\over (\vb{L}_2 -\vb{p})^2 -\vb{p}^2 -\iu 0 } 
    \\
    &= \left[ {\iu Z\overline{\alpha} \over \vel} (- 4p^2/\mu^2 - \iu 0)^{-\epsilon} \right]^2 \left[ {1\over 8\epsilon^2} + {\pi^2\over 12} +  5\zeta(3)\epsilon + \Order(\epsilon^2) \right] \,,\
    \end{split}
\end{equation}
where the integral is evaluated in \cref{sec:integrals}. 
At third order,
\begin{equation}
    \begin{split}
    \label{eq:SC3loophard}
    {{\cal M}_H^{(3)}}
    &=
    (2mZe^2)^3 \int {\dd^\D L_1 \over (2\pi)^\D} \int {\dd^\D L_2 \over (2\pi)^\D} \int {\dd^\D L_3 \over (2\pi)^\D} 
    {1\over \vb{L}_1^2}{1\over (\vb{L}_1 -\vb{p})^2 -\vb{p}^2 -\iu 0} {1\over (\vb{L}_1 - \vb{L}_2)^2} 
    {1\over (\vb{L}_2 -\vb{p})^2 -\vb{p}^2 -\iu 0}  \times
    \\
    &\qquad \times
    {1\over (\vb{L}_2 - \vb{L}_3)^2} 
    {1\over (\vb{L}_3 -\vb{p})^2 -\vb{p}^2 -\iu 0} 
    \\
    &= \left[ {\iu Z\overline{\alpha} \over \vel} (- 4p^2/\mu^2 - \iu 0)^{-\epsilon} \right]^3
    \left[ {-1\over 48\epsilon^3} - {\pi^2\over 24 \epsilon} - {13\zeta(3)\over 6}  + \Order(\epsilon) \right] \,.
    \end{split}
\end{equation}
The evaluation of this integral is also performed in \cref{sec:integrals}. 
At higher-loop order, direct evaluation of integrals becomes increasingly difficult.  We will see how wavefunction methods provide a closed-form expression for arbitrary loop order.

\subsection{Renormalization}

Before turning to the all-orders discussion, we present the renormalized hard matching coefficient through
three-loop order in the $\overline{\rm MS}$ scheme.
Identifying the above amplitudes as bare matching coefficients,
${\cal M}_H \equiv {\cal M}_H^{\rm bare}$, writing 
\begin{align}
    {\cal M}_H^{\rm bare} = \mathcal{Z}^{-1} {\cal M}_H(\mu)  \,,
\end{align}
and requiring that $\mathcal{Z}(\mu)$ has only $1/\epsilon$ terms when expressed in terms of $\overline{\alpha}$, 
we find
\begin{align}
    \mathcal{Z}^{-1} = 1+ \sum_{n=1}^\infty \left( Z \overline{\alpha}\over \vel\right)^n z^{(n)} \,,
\end{align}
with 
\begin{align}\label{eq:SCzi}
    z^{(1)} = {-\iu\over 2\epsilon} \,, \quad
    z^{(2)} = {-1\over 8\epsilon^2} \,, \quad
    z^{(3)} = {\iu\over 48\epsilon^3} \,. 
\end{align}
The renormalized matching coefficient (at $\epsilon=0$) is then
\begin{equation}
    \begin{split}
    \label{MHmu_schro}
    {\cal M}_H(\mu) &= 1 + {Z \alpha \over \vel} 
    \bigg( {\pi \over 2} +\iu\log{2p\over\mu} \bigg) 
    + \left( {Z \alpha \over \vel} \right)^2 \left( {\pi^2 \over 24} + {\iu\pi \over 2}\log{2p\over\mu} -\frac12\log^2{2p\over\mu} \right)
    \\
    &\quad
    + \left( Z \alpha \over \vel\right)^3 
    \left( - {\pi^3 \over 48} - {\iu\zeta(3)\over 3} 
    + {\iu\pi^2\over 24} \log{2p\over\mu} 
    - {\pi \over 4} \log^2{2p\over \mu} 
    -{\iu\over 6} \log^3{2p\over \mu}
    \right) 
    + \Order(\alpha^4) \,, 
    \end{split}
\end{equation}
where $\overline{\alpha}$ reduces to the on-shell 
QED coupling $\alpha$ at $\epsilon \to 0$
(recall that there are no dynamical 
leptons in the non-relativistic theory). 
Since the product ${\cal M}_S {\cal M}_H$ is UV and IR finite (at $\lambda \ne 0$),
the quantity $\mathcal{Z}$ is identical to the ($\overline{\rm MS}$) operator
renormalization constant for the soft operator, 
\begin{align}
    {\cal M}_S^{\rm bare} = \mathcal{Z} {\cal M}_S(\mu) \,.
\end{align}
From the explicit form of \cref{eq:SCsoft,eq:SC1loopsoft}, 
the renormalization constant to all orders is given by  
\begin{align}
    \mathcal{Z} = \exp\left( {\iu Z\overline{\alpha}\over 2\vel\epsilon}\right) \,,
\end{align}
in agreement through three-loop order with \cref{eq:SCzi}. 
The renormalized soft function is 
\begin{align}\label{eq:ren_soft}
    {\cal M}_S(\mu) &= \exp\left( {\iu Z\alpha\over\vel} \log{\mu\over\lambda} \right)\,.
\end{align}
%

\subsection{Wavefunction solution and all-orders hard function}

We recognize \cref{eq:SCgeneral} as the perturbative 
expansion of the position-space wavefunction evaluated at $\vb{r}=0$ 
for a particle scattered by a Coulomb source and described by the 
Hamiltonian, 
\begin{align}
    H = {p^2\over 2m} -{Z\alpha\over r} \e^{-\lambda r} \,. 
\end{align}
The all-orders solution at leading power is (see \cref{app:wf-schr})\,,
\begin{align} \label{coul-schro-ans}
    {\cal M} &= [\psi^{(-)}(0)]^* = \Gamma\left( 1 - {\iu Z\alpha\over \vel} \right)
    \exp\bigg[ {Z\alpha\over \vel}\left( {\pi \over 2}
    +\iu\log{2p\over\lambda} -\iu\gamma_{\rm E} \right) \bigg] + \Order\qty(\frac{\lambda}{p})~,
\end{align}
where $\psi^{(-)}$ denotes the scattering solution that matches asymptotically to 
a plane wave plus an ingoing spherical wave.  
Combining \cref{eq:ren_soft,coul-schro-ans} we obtain the closed form result
\begin{align}
    \label{MH-schro}
    {\cal M}_H(\mu) &= 
    { {\cal M} \over {\cal M}_S(\mu) }
    = \Gamma\left( 1 - {\iu Z\alpha\over \vel} \right)
    \exp\bigg[ {Z\alpha\over \vel}\left( {\pi \over 2}
    +\iu\log{2p\over\mu} -\iu\gamma_{\rm E} \right) \bigg] \,.
\end{align}
This result reproduces the above results, {\it cf.} \cref{MHmu_schro}, through three-loop order. 

\section{Dirac-Coulomb problem \label{sec:dirac}}

In place of \cref{eq:SCgeneral}, 
consider the amplitudes for a relativistic fermion in the Coulomb field of an extended object with a charge form factor $F(\vb{L}^2)$
\begin{equation}
    \begin{split}\label{eq:DCgeneral}
    \bar{u}(p) {\cal M} &=  \sum_{n=0}^\infty(Z e^2)^n 
    \int {\dd^\D L_1 \over (2\pi)^\D} \int {\dd^\D L_2 \over (2\pi)^\D}\cdots  \int {\dd^\D L_n \over (2\pi)^\D} 
    \\
    &\qquad \times {F(\vb{L}_1^2)\over \vb{L}_1^2 + \lambda^2}{1\over (\vb{L}_1 -\vb{p})^2 -\vb{p}^2-\iu 0 } {F( (\vb{L}_1-\vb{L}_2)^2) \over (\vb{L}_1 - \vb{L}_2)^2 + \lambda^2} 
    {1\over (\vb{L}_2 -\vb{p})^2 -\vb{p}^2 -\iu 0} \cdots 
    \\
    &\qquad\qquad  \times 
     {F( (\vb{L}_{n-1}-\vb{L}_{n-2})^2) \over (\vb{L}_{n-1} - \vb{L}_{n})^2 + \lambda^2} 
    {1\over (\vb{L}_n -\vb{p})^2 -\vb{p}^2 -\iu 0} 
    \\
    &\qquad\qquad\qquad \times
    \bar{u}(p) \gamma^0 ( \slashed{p} - \slashed{L}_1 + m) \gamma^0 (\slashed{p} - \slashed{L}_2 + m)  \cdots \gamma^0 ( \slashed{p} - \slashed{L}_n + m) 
    \,.
    \end{split}
\end{equation}
The Dirac-Coulomb problem corresponds to the hierarchy $p\sim m \ll \Lambda_{\rm UV}$. For $F(\vb{L}^2)=1$, $E=m$, and $\slashed{p} - \slashed{L}_i + m \to 2m$, the amplitude reduces to
the Schrodinger Coulomb problem (\ref{eq:SCgeneral}).  The fermionic case represented by \cref{eq:DCgeneral} involves nontrivial Dirac structure, and a dependence on UV momentum scales
$|\vb{L}| \gg p$.  In the limit of a point-like source we have $F(\vb{L}^2)=1$.  Similar to the Schrodinger-Coulomb case, 
we first consider the low-order contributions.  At one-loop, for $\lambda\to 0$ and $\epsilon\to 0$,
\begin{equation}
    \begin{split}
    {\cal M}^{(1)} &= 2E Ze^2 \int {\dd^\D L \over (2\pi)^\D}{1\over \vb{L}^2 + \lambda^2}{1\over (\vb{L} -\vb{p})^2 -\vb{p}^2 -\iu 0 }
     \left[ 1 - {1\over 2E} \gamma^0 \slashed{L} \right] 
     \\
     &\to {\iu Z\overline{\alpha}\over \vel} \left[ \left( \log{2p\over\lambda} - {\iu\pi \over 2} \right) 
     + \frac12 \left( {m\gamma^0 \over E} - 1 \right) \right]  \,.
    \end{split}
\end{equation}
Similar to \cref{eq:SCfac}, we can express the result, up to $\lambda/E$ power corrections as the product of soft and hard factors,
with ${\cal M}_S$ as in \cref{eq:SC1loopsoft}, and 
${\cal M}_H$ now containing two different Dirac structures,  
\begin{align}
    {\cal M}_H &= {\cal M}_{H1} + \left( {m\gamma^0 \over E} - 1 \right) {\cal M}_{H2} \,. 
\end{align}
At tree level, the hard factor is given by
\begin{align}
 {\cal M}_{H1}^{(0)} = 1 \,, \qquad {\cal M}_{H2}^{(0)} =0 \,, 
\end{align}
and at one loop, 
\begin{align}\label{eq:DC1loophard}
    {\cal M}_{H1}^{(1)}
    &= \left[ {\iu Z\overline{\alpha} \over \vel} (- 4p^2/\mu^2 - \iu 0)^{-\epsilon} \right]\left[ -1\over 2\epsilon \right]
    \,, \quad
     {\cal M}_{H2}^{(1)}
    = \left[ {\iu Z\overline{\alpha} \over \vel} (- 4p^2/\mu^2 - \iu 0)^{-\epsilon} \right]\left[ 1\over 2(1-2\epsilon) \right] \,.
\end{align}
At two loop order, using integrals from \cref{sec:integrals}, 
\begin{align}\label{eq:MH2loop}
     {\cal M}_{H1}^{(2)}
    &= 
    \left[ {\iu Z\overline{\alpha} \over \vel} (- 4p^2/\mu^2 - \iu 0)^{-\epsilon} \right]^2 \left[ {1\over 8\epsilon^2} + {\pi^2\over 12} + \vel^2 \left( {-1\over 8\epsilon} - \frac54  \right) + \Order(\epsilon) \right]
    \,, \nl
     {\cal M}_{H2}^{(2)}
    &= \left[ {\iu Z\overline{\alpha} \over \vel} (- 4p^2/\mu^2 - \iu 0)^{-\epsilon} \right]^2\left[ {-1\over 4\epsilon} -\frac12 + \Order(\epsilon) \right] \,.
\end{align}
%

\subsection{Factorization \label{sec:dirac-factor}}

The integrals in  \cref{eq:DCgeneral} are
UV divergent by power counting when $F(\vb{L}^2)=1$. 
The explicit computations above show that  
${\cal M}_S {\cal M}_H$ is UV divergent beginning at two loop order, indicating sensitivity to short distance physics.  
Regulating UV divergences with $F(\vb{L}^2)$ introduces a new UV scale, and a corresponding momentum region 
in loop diagrams with $|\vb{L}| \sim \Lambda_{\rm UV} \gg p$.  
The factorization formula is 
\begin{align}\label{eq:DCfac}
    {\cal M} &= {\cal M}_{S} {\cal M}_H {\cal M}_{\rm UV} \,. 
\end{align}
In the following, we compute the explicit form 
of ${\cal M}_{\rm UV}$ using an illustrative charge form factor.  We then introduce an alternative finite-distance regulator that permits an all-orders solution of ${\cal M}_{\rm UV}$.  Combined with an all-orders solution for the total amplitude ${\cal M}$ using the same UV regulator, and the all-orders solution of ${\cal M}_S$, we then extract ${\cal M}_H$ to all orders in perturbation theory.   
%

\subsection{UV contribution from a charge form factor \label{sec:UV-FF}} 
In dimensional regularization, the factor ${\cal M}_{\rm UV}$ is computed by setting $\lambda=p=0$.  For simplicity, we take 
\begin{align}
    \label{model-FF}
    F(\vb{L}^2) &= {\Lambda_{\rm UV}^2 \over \Lambda_{\rm UV}^2 + \vb{L}^2} \,. 
\end{align}
At one loop order, 
\begin{align} \label{eq:DC1loopUV}
    {\cal M}_{\rm UV}^{(1)} &=  Z e^2 
    \int {\dd^\D L \over (2\pi)^\D} 
    {F(\vb{L}^2)\over (\vb{L}^2)^2 } 
   \gamma^0  \vb*{\gamma}\cdot \vb{L} = 0 \,. 
\end{align}
Nontrivial contributions begin at two-loop order, 
\begin{align} \label{eq:DC2loopUV}
    {\cal M}^{(2)}_{\rm UV} &=  (Z e^2)^2
    \int {\dd^\D L_1 \over (2\pi)^\D} \int {\dd^\D L_2 \over (2\pi)^\D}
    {F(\vb{L}_1^2)\over (\vb{L}_1^2)^2 }
    {F( (\vb{L}_1-\vb{L}_2)^2) \over \vb{L}_2^2 (\vb{L}_1 - \vb{L}_2)^2 }
   \gamma^0  \vb*{\gamma}\cdot \vb{L}_1
   \gamma^0  \vb*{\gamma}\cdot \vb{L}_2
   \nl
  &= \left[ Z\overline{\alpha}\left(\mu / \Lambda_{\rm UV} \right)^{2\epsilon} \right]^2 
   \left[ 
   -\frac{1}{8\epsilon} - \frac12 + \Order(\epsilon) 
   \right] 
     \,.
\end{align}
Let us compute renormalized expressions through two-loop order. 
In the $\overline{\rm MS}$ scheme, the renormalized soft function is again given by \cref{eq:ren_soft}, 
\begin{align}\label{eq:DCS2}
    {\cal M}_S(\mu_S) &= 1 + {\iu Z \alpha \over \vel} \log{\mu_S\over\lambda} 
    - {(Z \alpha)^2 \over 2\vel^2} \log^2{\mu_S\over\lambda} 
    + \Order(\alpha^3)\,.
\end{align}
The renormalized hard function through 
two loop order is 
\begin{equation}
    \begin{split}
    \label{eq:DCH2}
  {\cal M}_H(\mu_S,\mu_H) &= 1 + {Z \alpha \over \vel} 
    \bigg[\iu\left(\log{2p\over\mu_S} - {\iu\pi \over 2}\right)
    + {\iu\over 2} \left( {m\over E}\gamma^0 - 1 \right) \bigg] 
    + \left( {Z \alpha \over \vel} \right)^2 \bigg\{ {-\pi^2 \over 12} 
     -\frac12 \left( \log{2p\over\mu_S} - {\iu\pi \over 2}\right)^2
     \\
     &\quad
      - \frac12 \left( \log{2p\over\mu_S} - {\iu\pi \over 2}\right)\left( {m\over E}\gamma^0 - 1 \right)
   + \left[ \frac54 - \frac12\left( \log{2p\over\mu_H} - {\iu\pi \over 2}\right)\right]\vel^2
   \bigg\}
    + \Order(\alpha^3) \,,
    \end{split}
\end{equation}
and the renormalized UV function for the form factor in \cref{model-FF} is 
\begin{align}\label{eq:DCUV2}
    {\cal M}_{\rm UV}(\mu_H) &= 1 + \left( {Z \alpha} \right)^2 \bigg[ - \frac12 - \frac12 \log{\mu_H\over\Lambda_{\rm UV}} \bigg] + \Order(\alpha^3)\,.
\end{align}
It is readily checked that with the explicit results (\ref{eq:DCS2}), (\ref{eq:DCH2}) and 
(\ref{eq:DCUV2}), 
the product (\ref{eq:DCfac}) is independent of $\mu_S$ and $\mu_H$ through two loop order. 

\subsection{UV contribution with finite distance regulator \label{sec:x-reg}}

Consider the series of amplitudes representing the perturbative expansion of the Dirac wavefunction at finite distance: 
\begin{equation}
    \begin{split}\label{eq:DC-xreg}
    \bar{u}(\vb{p}) 
    {\cal M}_{\vb{r}}
    &=  \sum_{n=0}^\infty(Z e^2)^n 
    \int {\dd^\D L_1 \over (2\pi)^\D} \int {\dd^\D L_2 \over (2\pi)^\D}\cdots  \int {\dd^\D L_n \over (2\pi)^\D} 
    \e^{-\iu \vb{L}_n\cdot \vb{r} } {1\over \vb{L}_1^2 + \lambda^2}{1\over (\vb{L}_1 -\vb{p})^2 -\vb{p}^2 -\iu 0}
    \\
    &\qquad \qquad \qquad \times  {1 \over (\vb{L}_1 - \vb{L}_2)^2 + \lambda^2} 
    {1\over (\vb{L}_2 -\vb{p})^2 -\vb{p}^2 -\iu 0} \cdots
     {1 \over (\vb{L}_{n-1} - \vb{L}_{n})^2 + \lambda^2} 
    {1\over (\vb{L}_n -\vb{p})^2 -\vb{p}^2 -\iu 0} 
    \\
    &\qquad \qquad \qquad \qquad \qquad \qquad \times
    \bar{u}(p) \gamma^0 ( \slashed{p} - \slashed{L}_1 + m) \gamma^0 (\slashed{p} - \slashed{L}_2 + m)  \cdots \gamma^0 ( \slashed{p} - \slashed{L}_n + m) 
    \,.
    \end{split}
\end{equation}
For loop momentum $|\vb{L}| \gg 1/|\vb{r}|$ the rapid oscillations of the exponential regulate the integral, and  the finite distance $r$ acts as UV regulator.  In the limit $1/r \gg p$, the amplitudes are described by the factorization theorem \cref{eq:DCfac}.

 The finite distance regulator 
 is convenient since regulated amplitudes correspond to coordinate space solutions of the Dirac equation,  which for $|\vb{p}|\ll 1/r$  have a closed form  solution ({\it cf.} \cref{app:wf-dirac}).  
 We may relate the finite distance regulator scheme to a conventional $\overline{\rm MS}$-regulated amplitude by  applying the method of regions \cite{Beneke:1997zp,Jantzen:2011nz}. The finite-distance regulated amplitude ${\cal M}_{\vb{r}}$ satisfies the factorization theorem (\ref{eq:DCfac}),
 \begin{equation}
    \label{eq:M-r}
     \mathcal{M}_{\vb{r}} = \mathcal{M}_S \mathcal{M}_H \mathcal{M}_{\rm UV} (\vb{r})~,
 \end{equation}
 where  the UV matching coefficient depends on $\vb{r}$.

We will now show that $\mathcal{M}_{\rm UV}(\vb{r})$ can be computed to all orders in perturbation theory. This fact is related to the structure of the loop integrals with a finite distance regulator, \cref{eq:DC-xreg}, as compared to a charge form factor, \cref{eq:DCgeneral}: the regulator affects only the final ( $\dd^\D L_n$ ) loop integration, so that all of the preceding integrals are recursively one-loop. Details are presented in \cref{app:all-orders-xreg}, with the
 results for  bare amplitudes at arbitrary even and odd orders in perturbation theory respectively:
\begin{align}
    \mathcal{M}_{\rm UV}^{(2n)} &= 
        \frac{(-1)^n}{n!} \qty(\frac{(Z\widetilde{\alpha})^2/8}{\epsilon})^n  ~\qty[\prod_{m=0}^{n-1} \frac{1}{ 1+2 m \epsilon }]~,
        \label{bare-even}\\
       \mathcal{M}_{\rm UV}^{(2n+1)}&=  \frac{(-1)^n}{n!} \qty(\frac{(Z\widetilde{\alpha})^2/8}{\epsilon})^n  ~\qty[\prod_{m=0}^{n} \frac{1}{ 1+2 m \epsilon }]\times 
       \qty[ - Z\widetilde{\alpha} \frac{\iu \gamma_0\vb*{\gamma}\cdot \hat{\vb{r}}}{2}
       ]\,. \label{bare-odd}
\end{align}
The quantity $\widetilde{\alpha}$ is given in terms of the $\overline{\rm MS}$ coupling 
$\overline{\alpha}$ in \cref{eq:msbar}, by
\begin{align}
    \widetilde{\alpha} \equiv \overline{\alpha} \times \left(\mu^2 r^2 \over 16\right)^{\epsilon} {\Gamma\left(\frac12-\epsilon\right)\over\Gamma\left(\frac12+\epsilon\right)} \,.
\end{align}
As discussed in \cref{app:all-orders-xreg}, both series can be expressed in closed form for arbitrary nonzero $\epsilon$ in terms of Bessel functions.  
The $\overline{\rm MS}$ renormalization constant can also be computed in closed form. A careful treatment of the small-$\epsilon$ asymptotics of the bare amplitudes, {\it cf.} \cref{app:all-orders-xreg}, then yields the all-orders result,
\begin{equation}\label{UV-func}
    \mathcal{M}_{\rm UV}(\mu) = \qty(\mu r ~\e^{\gamma_{\rm E}})^{\gammatrad -1}\frac{1+\gammatrad }{2\sqrt{\gammatrad }} \qty[ 1 - \frac{Z\alpha}{1+\gammatrad } \iu \gamma_0 \vb*{\gamma}\cdot \hat{\vb{r}}]~,
\end{equation}
where $\gammatrad= \sqrt{1-(Z\alpha)^2}$. The result (\ref{UV-func}) is renormalized in the $\overline{\rm MS}$ scheme using the coupling defined in \cref{eq:msbar}. 
%

\subsection{Wavefunction solution and all-orders hard function \label{sec:wf-dirac-all-orders}} 

The amplitude (\ref{eq:DC-xreg}) is related to the perturbative expansion of a solution to the Dirac equation, 
\begin{align}
    \left( -\iu \gamma^0 \vb*{\gamma}\cdot\vb*{\partial} + m\gamma^0 - {Z\alpha\over r}\e^{-\lambda r} \right) \psi = E\psi \,,
\end{align}
namely: 
\begin{align}\label{eq:Mfrompsi}
    \bar{u}(p){\cal M} = [\psi^{(-)}(-\bm{r})]^\dagger \gamma^0 \,,
\end{align}
where $\psi^{(-)}(\bm{r})$ denotes the solution that is 
asymptotically a plane wave plus incoming spherical wave. 
The solution, ignoring power corrections in $\lambda/p$ and $p/r^{-1}$, is 
\begin{align}
\label{jackson-wf-fermifunction}
    \psi^{(-)}(\bm{r}) 
    &= \e^{\iu\phi} \sqrt{ E + \gammatrad m \over E + m} \sqrt{ F(Z,E,r)} \bigg( 1 + {\iu Z\alpha \over 1+\gammatrad }\gamma^0\vb*{\gamma}\cdot\hat{\vb*{r}} \bigg)
    \bigg[ {1+M\over 2} + {1-M\over 2}\gamma^0 \bigg]u(\bm{p})  \,.
\end{align}
Here $F(Z,E,r)$ is the Fermi function,
\begin{align}
    F(Z,E,r) = {2 (1+\gammatrad) \over [\Gamma(2\gammatrad +1)]^2} |\Gamma(\gammatrad +\iu\xi)|^2 \e^{\pi \xi} (2pr)^{2(\gammatrad -1)} \,,
\end{align}
the phase factor $e^{\iu\phi}$ is given by 
\begin{align}
    \e^{\iu\phi} &= 
    \e^{-\iu\xi \left(\log{2p\over\lambda} - \gamma_{\rm E}\right) +\iu(\gammatrad -1){\pi \over 2}} {\Gamma(\gammatrad +\iu \xi)\over |\Gamma(\gammatrad +\iu\xi)|}
    \sqrt{ \gammatrad + \iu \xi \over 1 +\iu\xi {m\over E} } \,,
\end{align}
and the quantity $M$ is given by
\begin{align}
    M &= {E+m\over E+\gammatrad m} \left( 1 +\iu\xi {m\over E}\right) \,.    
\end{align}
In the Dirac (i.e., ``Bjorken and Drell'') basis for $\gamma^\mu$ and with relativistic normalization $u(\bm{p})^\dagger u(\bm{p}) = 2E$, 
the expression is 
\begin{align}
      \psi^{(-)}(\bm{r}) 
    &= \e^{\iu\phi} \sqrt{ F(Z,E,r)} \bigg( 1 + {\iu Z\alpha \over 1+\gammatrad }\gamma^0\vb*{\gamma}\cdot\hat{\vb*{r}} \bigg)
    U(\bm{p})  \,,
\end{align}
where
\begin{align}
    U(\bm{p}) &= \sqrt{E+\gammatrad m} 
    \left(\begin{array}{c} 1 \\ \\  \left( 1 +\iu\xi {m\over E}\right)  { \bm{\sigma}\cdot \bm{p} \over E+\gammatrad m} 
    \end{array}\right) \chi \,, 
\end{align}
and $\chi$ is a two-component spinor. 
Using \cref{eq:Mfrompsi}, the explicit all-orders results for ${\cal M}_S$ in \cref{eq:ren_soft}, and ${\cal M}_{\rm UV}$ in \cref{UV-func}, 
the hard function appearing in the factorization formula (\ref{eq:DCfac}) is 
\begin{equation}
    \begin{split}\label{eq:MHDren}
    \mathcal{M}_H(\mu_S,\mu_H) &= \mathcal{M}_S^{-1}(\mu_S){\cal M} \mathcal{M}_{\rm UV}^{-1}(\mu_H) 
    \\
    &= \e^{ {\pi \xi \over 2} + \iu\xi \left(\log{2p\over\mu_S} - \gamma_{\rm E}\right) - \iu(\gammatrad -1){\pi \over 2}} {2 \Gamma(\gammatrad -\iu \xi) \over \Gamma(2\gammatrad +1)}
    \sqrt{ \gammatrad - \iu \xi \over 1 -\iu\xi {m\over E} }  
   \sqrt{ E + \gammatrad m \over E + m} 
   {\sqrt{2 \gammatrad \over 1+\gammatrad }}
   \left( 2p\e^{-\gamma_{\rm E}}\over \mu_H\right)^{\gammatrad -1} 
   \\
   &\hspace{0.55\linewidth} 
   \times
   \bigg[ {1+M^*\over 2} + {1-M^*\over 2}\gamma^0 \bigg]
   \,.
    \end{split}
\end{equation}
The amplitude has been explicitly decomposed into separate factors depending on a single scale, $\lambda$, $p$, or $r^{-1}$ (here we are not distinguishing the scales $p$, $m$, and $E$). We remark that the explicit appearance of $\exp(\gamma_{\rm E})$ accompanying $2p/\mu$
in \cref{eq:MHDren} may seem unexpected since the hard amplitude must match 
conventional $\overline{\rm MS}$ renormalized amplitudes order by order in perturbation theory.  
However, these factors cancel against implicit factors\footnote{
This can be seen most easily by noting that the two perturbative parameters that appear are $\gammatrad -1\sim {\cal O}([Z\alpha]^2)$ and $\xi \sim {\cal O}(Z\alpha)$. Then, using $\Gamma(1+2\gammatrad)=  
2\gammatrad (2\gammatrad -1) \Gamma(1+2(\gammatrad -1))$ and
    $\log \Gamma(1+z) = -\log(1+z) + z(1-\gamma_{\rm E}) + \sum_{n=0}^\infty (-1)^n \qty(\zeta(n)-1)\frac{z^n}{n}$, 
it is easy to show that the combination 
    $\e^{-\iu \xi \gamma_{\rm E}} \Gamma(\gammatrad -\iu\xi) \e^{-(\gammatrad -1)\gamma_{\rm E}}/ \Gamma(1+2\gammatrad)$
contains no factors of $\gamma_{\rm E}$ at any order in perturbation theory. 
} of $\gamma_{\rm E}$ from the expansion of $\Gamma(\gammatrad -\iu \xi)/\Gamma(1+2\gammatrad)$.

Given \cref{eq:MHDren} we can extract the anomalous dimension for contact operators to all orders in $Z\alpha$. We differentiate $\mathcal{M}_H$ with respect to $\mu_H$ and obtain
\begin{equation}
    \label{anom-dim}
    \gamma_{\mathcal{O}} = \sqrt{1-(Z\alpha)^2} -1 \,. 
\end{equation}
This is the contribution to the anomalous dimension from each light-particle leg. For example the operator mediating \cref{eq:AB} has an anomalous dimension of  $2 \gamma_\mathcal{O}$.  
For an operator mediating a beta decay, $A[Z+1]\rightarrow B[Z] + \ell^+ \nu$, \cref{anom-dim} is the leading-$Z$ contribution to the anomalous dimension \cite{Hill:2023acw,z2a3anom}. Including the one-loop beta function with $n_f$ dynamical fermions, the scale dependence of contact operators can be obtained in closed form 
\begin{equation}
     \label{eq:RG-sol}
    \int_{\alpha_L}^{\alpha_H} \dd \alpha' ~\frac{\gamma_H(\alpha')}{\beta(\alpha')}
    =\int_{\alpha_L}^{\alpha_H} \dd \alpha'~ \frac{\sqrt{1-Z^2\alpha^{\prime 2}}-1}{\tfrac{2n_f}{3\pi} \alpha^{\prime 2}} = \frac{3\pi}{2n_f}
     \qty{\frac{1-\gammatrad_H }{\alpha_H} - \frac{1-\gammatrad_L}{\alpha_L} - Z \qty[\arcsin(Z\alpha_H)-\arcsin(Z\alpha_L)]}~,
\end{equation}
where we have introduced the notation $\gammatrad_{L,H}=\gammatrad(\alpha_{L,H})$. This expression is useful when analyzing QED radiative corrections for the beta decays of heavy nuclei \cite{Hill:2023acw}. 

The hard function (\ref{eq:MHDren}), 
describes the limit $p\sim m \ll \Lambda_{\rm UV}$, where 
$\Lambda_{\rm UV} ~\sim R^{-1}$ 
denotes the scale of nuclear or hadronic structure. When the lepton is non-relativistic, $p \ll m \ll \Lambda_{\rm UV}$, it is convenient to expand the hard function as 
\begin{align}
    {\cal M}_H = {\cal M}_{H}^+ P_+ + {\cal M}_{H}^- P_- \,,
\end{align}
where $P_\pm = (1\pm \gamma^0)/2$.  Allowing for arbitrary values of $\xi$, we find through second order in $\beta$, 
\begin{equation}
    \begin{split}
    \label{non-rel}
    \mathcal{M}_{H}^+  &= \e^{{\pi \xi \over 2} 
    +\iu \xi \left(\log{2p\over\mu_S} - \gamma_{\rm E}\right) } \Gamma(1-\iu \xi) 
    \bigg\{ 1 + \beta^2\bigg[- {\iu\over 4}\xi     
    + \xi^2 \left( -\frac12 \log{2p\over\mu_H}+\frac54+{\iu \pi \over 4} -{\gamma_{\rm E}\over 2} -\frac12\psi(1-\iu \xi)  \right) \bigg]\bigg\}
   \,, 
\\
\mathcal{M}_{H}^- &= \e^{{\pi \xi \over 2} 
    +\iu \xi \left(\log{2p\over\mu_S} - \gamma_{\rm E}\right) } \Gamma(2-\iu \xi) 
    \bigg\{ 1 + \beta^2\bigg[ {\iu\over 4}\xi     
    + \xi^2 \left( -\frac12 \log{2p\over\mu_H}+ \frac32 +{\iu \pi \over 4} -{\gamma_{\rm E}\over 2} -\frac12\psi(2-\iu \xi)  \right) \bigg]\bigg\}
   \,, 
   \end{split}
\end{equation}
where $\psi$ denotes the digamma function, $\psi(x) = \Gamma^\prime(x)/\Gamma(x)$. 
At each order in $\beta^2$, the expressions (\ref{non-rel}) 
sum an infinite series of terms involving powers $\xi^n$. 
At $\beta\to 0$, the leading term for the 
``large'' upper component ${\cal M}_H^+$ reduces to the Schrodinger-Coulomb result (\ref{MH-schro}) which corresponds to a ``non-relativistic Fermi function'', {\it cf.} Refs.~\cite{Wilkinson:1982hu,Cirigliano:2023fnz}.  

\section{Discussion \label{sec:discuss}}

The formula (\ref{eq:DCfac}), and its non-relativistic analog (\ref{eq:SCfac}), provides an all-orders explicit demonstration of factorization for the Coulomb problem.  We find that Coulomb corrections factorize among different legs for a contact interaction (see \cref{sec:coulomb-problem}). The universal hard matching coefficient in this formula, ${\cal M}_H$ in \cref{eq:MHDren}, can be applied to different processes, and large logarithms can be summed to all orders using renormalization group methods.  The non-relativistic limit for $p\ll m \ll \Lambda_{\rm UV}$ is given by \cref{non-rel}. By identifying the amplitudes as quantum field theory objects in a standard regularization scheme 
(i.e., $\overline{\rm MS}$ scheme in dimensional regularization), we can systematically compute subleading perturbative contributions and match to hadronic and nuclear matrix elements. More detailed discussions of these points are presented elsewhere~\cite{z2a3anom,Hill:2023acw,largepi}. It is interesting to note that for unpolarized observables to beta decay, the spin-summed matrix element squared,\!\footnote{Explicitly we define $\big\langle  |{\cal M}_H|^2 \big\rangle:=  \sum_{\rm spins} \left|\bar{u}\mathcal{M}_H\gamma_0 P_L v \right|^2  \Big/ \sum_{\rm spins} \left|\bar{u}\gamma_0 P_L v \right|^2$ where $\gamma_0 P_L = \gamma_\mu v^\mu P_L$ is the tree-level Dirac structure, with $v_\mu=(1,0,0,0)$.}
\begin{align}
    \label{FF-factorized}
    \big\langle  |{\cal M}_H|^2 \big\rangle 
     = F(Z,E)\big|_{r_H} \times \frac{4\gammatrad}{(1+\gammatrad)^2}
    \,,
\end{align}
differs from the historically defined Fermi function even when evaluated at $r_H^{-1}= \mu_H e^{\gamma_{\rm E}}$. We observe that finite-distance regulated amplitudes have special algebraic properties that allow for 
explicit all-orders expressions, for both bare and renormalized matrix elements as shown explicitly in \cref{bare-even,bare-odd,UV-func}. This example of all-orders renormalization may be of formal interest. 

As an illustration of how the formalism applies to different processes, 
let us return to \cref{eq:AB}.  For definiteness, suppose that the neutral current 
reaction is mediated by exchange of a vector boson of mass $m_{B}$.
The tree-level amplitude depicted in 
\cref{equality-1} takes the form 
\begin{align}
    {\cal M}^{\rm tree} = 
    {m_B^2 \over m_B^2 - (p_1+p_2)^2}  \bar{u}(\bm{p}_1) \Gamma^{\rm tree} v(\bm{p}_2) \,,
\end{align}
where $\Gamma^{\rm tree} = \gamma^0 (A+B\gamma_5)$ for some numbers $A$ and $B$.  When $\Lambda \gg m_B \gg p$, the boson mass plays the role of UV regulator.\!\footnote{
We have in mind a $Z^\prime$ boson extending the Standard Model.  The amplitudes are equivalent to a Standard Model $Z$ boson in the formal limit $\Lambda_{\rm nuc} \gg m_Z \gg m_e$.
}
The factorization formula describing the infinite sum in \cref{equality-1}, is
\begin{align}
    {\cal M} &= 
     \bar{u}(\bm{p}_1) 
     {\cal M}_S(\bm{p}_1) 
      {\cal M}_H(\bm{p}_1) 
      {\cal M}_{\rm UV}
      ~\overline{\!\!{\cal M}}_H(\bm{p}_2)
      ~\overline{\!\!{\cal M}}_S(\bm{p}_2)
     v(\bm{p}_2) \,.
\end{align}
Here the conjugate amplitude is denoted
$~~\overline{\!\!\cal M} = \gamma^0 {\cal M}^\dagger \gamma^0$.  
It is straightforward to compute ${\cal M}_{\rm UV}$ from the diagrams in \cref{equality-1}, neglecting charged lepton masses and momenta.  Through two-loop order, after $\overline{\rm MS}$ renormalization, 
\begin{align}
    {\cal M}_{\rm UV}(\mu) &= 
    \Gamma^{\rm tree} 
    \left[ 1 + (Z\alpha)^2 \left(\frac12 \log{\mu^2\over m_B^2}- \frac34 \right)
    \right]
    \,.
\end{align}
It is readily seen that the scale dependence of ${\cal M}_{\rm UV}(\mu_H)$ cancels against the product of ${\cal M}_H(\mu_H)$ for the charged leptons.

An important application of the formalism presented above is to the
description of precision nuclear beta decay, e.g., for $|V_{ud}|$ determination~\cite{Hardy:2020qwl} and tests of first row CKM unitarity \cite{Czarnecki:2004cw,Seng:2018yzq,Seng:2018qru,Czarnecki:2019iwz,Czarnecki:2019mwq,Coutinho:2019aiy,Hardy:2020qwl,Crivellin:2020ebi,Crivellin:2021njn,Crivellin:2020lzu,Cirigliano:2022yyo}. 
Consider the decay of a heavy atom to a negatively charged ion, a positron, and a neutrino~\cite{UCNA:2017obv,Darius:2017arh,Fry:2018kvq,UCNt:2021pcg,Shidling:2014ura,Eibach:2015ksa,Sternberg:2015nnr,Gulyuz:2016ppg,Fenker:2016mka,Long:2017gdh,Fenker:2017rcx,Brodeur:2016cci,Shidling:2018fvb,Valverde:2018haz,OMalley:2020vop,Burdette:2020bke,Long:2020lby,Muller:2022jew,Long:2022yea},
\begin{equation}
  ~A \rightarrow ~I^- + e^+ + \nu_e\,.\label{beta-decay-ion}
\end{equation}
Beta decays are a complicated multi-scale problem, 
involving energies from the weak scale $ \sim 100~{\rm GeV}$, down to scales set by atomic screening $\sim 100 ~{\rm eV}$. Structure dependent corrections e.g., due to nuclear charge distributions, can be subsumed into a short-distance Wilson coefficient in the point-like theory considered here. The embedding of Coulomb corrections in a broader EFT framework is crucial for the systematic separation of physical scales and computation of QED radiative corrections. For charged current processes such as beta decays, the charge-mismatch between the initial and final heavy particle (i.e., nucleus) introduce sub-leading effects whose analysis can be substantially simplified using eikonal algebra \cite{eikonal_algebra}. Systematic evaluation of these subleading corrections differ from previous phenomenological approaches and lead to numerical differences that are larger than the existing estimated error budget for outer radiative corrections \cite{Hardy:2020qwl,Hill:2023acw}; detailed calculations are presented elsewhere \cite{Hill:2023acw,z2a3anom}. 

\vskip 0.1in
\noindent{\bf Acknowledgements}
\vskip 0.1in
\noindent We thank Susan Gardner for useful discussions, and the Neutrino Theory Network for sponsoring RP's visit to U.\ Kentucky in 2018. RP thanks Benoit Assi, Florian Herren, and Robert Szafron for helpful discussions. This work was supported by the U.S. Department of Energy, Office of Science, Office of High Energy Physics, under Awards DE-SC0019095.  Fermilab is operated by Fermi Research Alliance, LLC under Contract No. DE-AC02-07CH11359 with the United States Department of Energy. Part of this research was performed at the Kavli Institute for Theoretical Physics which is supported in part by the National Science Foundation under Grant No. NSF PHY-1748958 and at the Aspen Center for Physics, which is supported by National Science Foundation grant PHY-1607611. This work is supported by the U.S. Department of Energy, Office of Science, Office of High Energy Physics, under Award Number DE-SC0011632 and by the Walter Burke Institute for Theoretical Physics. RP acknowledges support from the U.S. Department of Energy, Office of Science, Office of High Energy Physics, under Award Number DE-SC0011632 and the Neutrino Theory Network Program Grant under Award Number DE-AC02-07CHI11359.

\appendix 

\section{Loop integrals \label{sec:integrals}}

We collect here some results for loop integrals that are used in the main text.  Integrals are defined
in Euclidean $D$-dimensional space with $D=3-2\epsilon$.

\subsection{Two loop integrals}

\subsubsection{Scalar Integrals}
Consider the two-loop integral 
\begin{align}
    J(a_1,a_2,a_3,a_4,a_5)
    &= \int {\dd^\D L_2 \over (2\pi)^\D}  {\dd^\D L_1 \over (2\pi)^\D} 
    {1\over [ \vb{L}_2^2 ]^{a_1}}
    {1\over [ (\vb{p}-\vb{L}_2)^2-\vb{p}^2]^{a_2} }
    {1\over [ \vb{L}_1^2 ]^{a_3}} 
    {1\over [ (\vb{p}-\vb{L}_1)^2-\vb{p}^2]^{a_4} }
    {1\over [ (\vb{L}_1-\vb{L}_2)^2 ]^{a_5} }\,.
\end{align}
Using that the integral of a total derivative vanishes in dimensional regularization, 
and inserting $(\partial /\partial L_2^i) L_2^i$ and $(\partial /\partial L_2^i) L_1^i$ under the integral, yields the following ``integration by parts'' \cite{Smirnov:2004ym} relation, 
\begin{align}
    \label{IBP-tri}
    0 &= \D - a_1 - a_2 - 2a_5 - a_1 \bm{1}^+ (\bm{5}^- - \bm{3}^-)  - a_2 \bm{2}^+ ( \bm{5}^- - \bm{4}^- ) \,,
\end{align}
where we use the shorthand $\bm{m}^\pm$ to denote the raising or lowering indices
in $J$, e.g., $ \bm{2}^\pm J(a_1,a_2,a_3,a_4,a_5) = J(a_1,a_2\pm 1, a_3, a_4, a_5)$. 
In particular, for the two-loop integral appearing in \cref{eq:SC2loophard}, 
\begin{align}
    J(0,1,1,1,1) &= {1\over \D-3}\left[ J(0,2,1,1,0) - J(0,2,1,0,1) \right] \,,
\end{align}
where the integrals on the right-hand side are recursively one-loop and 
are readily evaluated: 
\begin{align}
    J(0,2,1,1,0) &= (-p^2-\iu 0)^{-1-2\epsilon}\left[ \Gamma\left(\frac12 + \epsilon\right) \over (4\pi)^{\frac32 -\epsilon} \right]^2\left( -1\over 2\epsilon \right) \label{master-integral-1}~ ,\\
    J(0,2,1,0,1) &= (-p^2-\iu 0)^{-1-2\epsilon}\left[ \Gamma\left(\frac12 + \epsilon\right) \over (4\pi)^{\frac32 -\epsilon} \right]^2 { \Gamma\left(\frac12-\epsilon\right)^2 \Gamma(1+2\epsilon) \Gamma(-4\epsilon) \over \Gamma\left(\frac12+\epsilon\right) \Gamma(1-2\epsilon)\Gamma\left(\frac12-3\epsilon\right)}\label{master-integral-2}\,.
\end{align}
%
\subsubsection{Vector and Tensor Integrals}
In the evaluation of the hard function for a relativistic lepton, \cref{eq:MH2loop}, we encounter the following two loop integrals,
\begin{align}
    J^i(a_1,a_2,a_3,a_4,a_5)&= \int {\dd^\D L_2 \over (2\pi)^\D} {\dd^\D L_1 \over (2\pi)^\D} 
    ~~{ L_2^i  \over [ \vb{L}_2^2 ]^{a_1}}
    {1\over [ (\vb{p}-\vb{L}_2)^2-\vb{p}^2]^{a_2} }
    {1\over [ \vb{L}_1^2 ]^{a_3}} 
    {1\over [ (\vb{p}-\vb{L}_1)^2-\vb{p}^2]^{a_4} }
    {1\over [ (\vb{L}_1-\vb{L}_2)^2 ]^{a_5} }~,\\
    J^{ij}(a_1,a_2,a_3,a_4,a_5)&= \int {\dd^\D L_2 \over (2\pi)^\D} {\dd^\D L_1 \over (2\pi)^\D}  
    ~~{ {L_{2}^{i} L_{1}^{j} } \over [ \vb{L}_2^2 ]^{a_1}}
    {1\over [ (\vb{p}-\vb{L}_2)^2-\vb{p}^2]^{a_2} }
    {1\over [ \vb{L}_1^2 ]^{a_3}} 
    {1\over [ (\vb{p}-\vb{L}_1)^2-\vb{p}^2]^{a_4} }
    {1\over [ (\vb{L}_1-\vb{L}_2)^2 ]^{a_5} } \,.
\end{align}
In particular, we require the contractions $p^i J^i(1,1,0,1,1)$, $p^i J^i(0,1,1,1,1)$, and $\delta^{ij} J^{ij}(0,1,1,1,1)$, which by partial-fractioning can be written, 
\begin{align}
    2 p^i J^i(1,1,0,1,1) &= J(0,1,0,1,1)-J(0,1,1,0,1) \,,
    \nl
    2 p^i J^i(0,1,1,1,1) &= J(-1,1,1,1,1)-J(0,0,1,1,1) \,,
    \nl
    2 \delta^{ij} J^{ij}(0,1,1,1,1) &= J(0,1,0,1,1)+J(-1,1,1,1,1)-J(0,1,1,1,0) \,.
\end{align}
Applying the integration by parts identity (\ref{IBP-tri}) yields 
\begin{align}
    J(-1,1,1,1,1) &= {1\over \D-2}\big[ -J(0,1,1,1,0)+ J(0,1,0,1,1)+J(-1,2,1,1,0)-J(-1,2,1,0,1) \big] \,. 
\end{align}
The remaining integrals are recursively one-loop and are given by 
\begin{equation}
    \begin{split}
    J(-1,2,1,0,1) &= (-p^2-\iu 0)^{-2\epsilon}\left[ \Gamma\left(\frac12 + \epsilon\right) \over (4\pi)^{\frac32 -\epsilon} \right]^2
    { \Gamma\left(\frac12-\epsilon\right)^2 \Gamma(2\epsilon) \Gamma(2-4\epsilon) \over \Gamma\left(\frac12+\epsilon\right) \Gamma(1-2\epsilon)\Gamma\left(\frac32-3\epsilon\right)} \,,
    \\
    J(-1,2,1,1,0) &= (-p^2-\iu 0)^{-2\epsilon}\left[ \Gamma\left(\frac12 + \epsilon\right) \over (4\pi)^{\frac32 -\epsilon} \right]^2
    { 2(1-\epsilon)
    \over \epsilon(1-2\epsilon) } \,,
    \\
    J(0,0,1,1,1) &= 0 \,,
    \\
    J(0,1,0,1,1) &= (-p^2-\iu 0)^{-2\epsilon}\left[ \Gamma\left(\frac12 + \epsilon\right) \over (4\pi)^{\frac32 -\epsilon} \right]^2
    {1\over \epsilon(1-2\epsilon)} \,,
    \\
   J(0,1,1,0,1) &= (-p^2-\iu 0)^{-2\epsilon}\left[ \Gamma\left(\frac12 + \epsilon\right) \over (4\pi)^{\frac32 -\epsilon} \right]^2
    { \Gamma\left(\frac12-\epsilon\right)^2 \Gamma(2\epsilon) \Gamma(1-4\epsilon) \over \Gamma\left(\frac12+\epsilon\right) \Gamma(1-2\epsilon)\Gamma\left(\frac32-3\epsilon\right)} \,,
    \\
    J(0,1,1,1,0) &= (-p^2-\iu 0)^{-2\epsilon}\left[ \Gamma\left(\frac12 + \epsilon\right) \over (4\pi)^{\frac32 -\epsilon} \right]^2
    {1\over \epsilon(1-2\epsilon)} \,.
    \end{split}
\end{equation}
%

\subsection{Three loop integrals}

Consider the three-loop integral,
\begin{equation}
    \begin{split}
    I(a_2,a_4,a_5)
    &= 
    \int {\dd^\D L_1 \over (2\pi)^\D}\!\! \int {\dd^\D L_2 \over (2\pi)^\D}\!\! \int {\dd^\D L_3 \over (2\pi)^\D} \\
    &\hspace{0.1\linewidth}
    {1\over \vb{L}_1^2}{1\over (\vb{L}_1 -\vb{p})^2 -\vb{p}^2 } {1\over (\vb{L}_1 - \vb{L}_2)^2} 
    {1\over [(\vb{L}_2 -\vb{p})^2 -\vb{p}^2]^{a_4} } 
    {1\over [(\vb{L}_2 - \vb{L}_3)^2]^{a_5} } 
    {1\over [(\vb{L}_3 -\vb{p})^2 -\vb{p}^2]^{a_2} } \,.
    \end{split}
\end{equation}
Integration by parts identities are ({\it cf.} \cref{IBP-tri} at $a_1=0$), 
\begin{align}
    0 &= \D - a_2 - 2a_5 - a_2 \bm{2}^+ ( \bm{5}^- - \bm{4}^- ) \,, 
\end{align}
so that the integral of interest in \cref{eq:SC3loophard} is 
\begin{align}\label{eq:Is}
    I(1,1,1) &= {1\over \D-3} \left[ I(2,1,0) - I(2,0,1) \right]\,.
\end{align}
The first integral in \cref{eq:Is} is given by the product of two- and one-loop integrals, 
\begin{equation}
    \begin{split}
    I(2,1,0) &=  
    \left[ \int {\dd^\D L_1 \over (2\pi)^\D}\!\! \int {\dd^\D L_2 \over (2\pi)^\D} 
     {1\over \vb{L}_1^2}{1\over (\vb{L}_1 -\vb{p})^2 -\vb{p}^2 } {1\over (\vb{L}_1 - \vb{L}_2)^2} 
    {1\over (\vb{L}_2 -\vb{p})^2 -\vb{p}^2 } 
    \right]
    \left[ \int {\dd^\D L_3 \over (2\pi)^\D} {1\over [(\vb{L}_3 -\vb{p})^2 -\vb{p}^2]^{2} } \right]
    \\
    &= J(0,1,1,1,1) (-p^2-\iu 0)^{-\frac12-\epsilon} {\Gamma\left(\frac12 + \epsilon\right)  \over (4\pi)^{\frac32 -\epsilon} } \,,
    \end{split}
\end{equation}
where $J(0,1,1,1,1)$ is evaluated above. 
The second integral in \cref{eq:Is} is recursively two-loop,
\begin{equation}
    \begin{split}
        \label{eq:I201}
    I(2,0,1) &=  
    \int {\dd^\D L_1 \over (2\pi)^\D}\!\! \int {\dd^\D L_3 \over (2\pi)^\D} 
     {1\over \vb{L}_1^2}{1\over (\vb{L}_1 -\vb{p})^2 -\vb{p}^2 } 
      {1\over [(\vb{L}_3 -\vb{p})^2 -\vb{p}^2]^{2} }
    \left[ \int {\dd^\D L_2 \over (2\pi)^\D} {1\over (\vb{L}_1 - \vb{L}_2)^2} 
    {1\over (\vb{L}_2 - \vb{L}_3)^2} 
    \right]
    \\
    &= \int {\dd^\D L_1 \over (2\pi)^\D}\!\! \int {\dd^\D L_3 \over (2\pi)^\D}
     {1\over \vb{L}_1^2}{1\over (\vb{L}_1 -\vb{p})^2 -\vb{p}^2 } 
      {[(\vb{L}_1-\vb{L}_3)^2]^{-\frac12-\epsilon} \over [(\vb{L}_3 -\vb{p})^2 -\vb{p}^2]^{2} }\times 
    {\Gamma\left(\frac12 + \epsilon\right)  \over (4\pi)^{\frac32 -\epsilon} } \betaFunc\left(\frac12-\epsilon, \frac12-\epsilon \right)
    \\
    &= {\Gamma\left(\frac12 + \epsilon\right)  \over (4\pi)^{\frac32 -\epsilon} }
    \betaFunc\left(\frac12-\epsilon, \frac12-\epsilon \right)
    J\left(0,2,1,1, \frac12 + \epsilon\right)
    \,.
    \end{split}
\end{equation}
To evaluate $J\left(0,2,1,1, \frac12 + \epsilon\right)$, we first perform the 
$\vb{L}_3$ integral in \cref{eq:I201}, 
\begin{align}
    \int {\dd^\D L_3 \over (2\pi)^\D}  {1\over [(\vb{L}_3 -\vb{p})^2 -\vb{p}^2]^{2} } [(\vb{L}_1-\vb{L}_3)^2]^{-\frac12-\epsilon} 
    = {\Gamma(1+2\epsilon) \over \Gamma\left(\frac12 + \epsilon\right) (4\pi)^{\D/2}} 
    \int_0^1 \dd x\, x^{-2\epsilon}(1-x)^{-\frac32 - \epsilon} \left[ (\vb{L}_1 - \vb{p})^2 - {\vb{p}^2\over 1-x} \right]^{-1-2\epsilon}  \,,
\end{align}
so that 
\begin{align}\label{eq:xint}
    J\left(0,2,1,1, \frac12 + \epsilon\right) 
    &=  {\Gamma(1+2\epsilon) \over \Gamma\left(\frac12 + \epsilon\right) (4\pi)^{\D/2}} 
    \int_0^1 \dd x\, x^{-2\epsilon}(1-x)^{-\frac32 - \epsilon}  K(1,1,1+2\epsilon) \,, 
\end{align}
where we introduce 
\begin{align}
    K(a_1,a_2,a_3) &= \int {\dd^\D L \over (2\pi)^\D} 
    {1\over \left[ \vb{L}^2\right]^{a_1} }
    {1\over \left[(\vb{L} -\vb{p})^2 -\vb{p}^2 \right]^{a_2}} 
    {1\over \left[ (\vb{L} - \vb{p})^2 - \vb{p}^2/ (1-x) \right]^{a_3} }\,.
\end{align}
Integration by parts for $K$ yields 
\begin{align}
    0 = \D - 2a_1 - a_2 - a_2 \bm{2}^+ \bm{1}^-   - a_3 \bm{3}^+ ( \bm{1}^- + \bm{2}^- ) \,,
\end{align}
so that 
\begin{align}\label{eq:Ks}
    K(1,1,1+2\epsilon) &= {1\over \D-3} \left\{ 
    K(0,2,1+2\epsilon) 
    + (1+2\epsilon)\left[ K(0,1,2+2\epsilon) + K(1,0,2+2\epsilon) \right] \right\} \,.
\end{align}
As a function of the integration variable $x$ in \cref{eq:xint}, the terms 
on the right side of \cref{eq:Ks} are
\begin{equation}
    \begin{split}
    K(0,2,1+2\epsilon) &= { (-p^2)^{-\frac32-3\epsilon} \over (4\pi)^{\D/2}}
    {\Gamma\left(\frac32 + 3\epsilon\right)\over \Gamma(1+2\epsilon)} 
    \int_0^1 \dd z\, z(1-z)^{2\epsilon} \left( z + {1-z\over 1-x} \right)^{-\frac32 - 3\epsilon} \,,
    \\
    K(0,1,2+2\epsilon) &= { (-p^2)^{-\frac32-3\epsilon} \over (4\pi)^{\D/2}}
    {\Gamma\left(\frac32 + 3\epsilon\right)\over \Gamma(2+2\epsilon)} 
     \int_0^1 \dd z\, (1-z)^{1+2\epsilon} \left( z + {1-z\over 1-x} \right)^{-\frac32 - 3\epsilon} \,,
     \\
      K(1,0,2+2\epsilon) &= { (-p^2)^{-\frac32-3\epsilon} \over (4\pi)^{\D/2}}
    {\Gamma\left(\frac32 + 3\epsilon\right)\over \Gamma(2+2\epsilon)} 
     \int_0^1 \dd z\, (1-z)^{1+2\epsilon} \left( -z(1-z) + {1-z\over 1-x} \right)^{-\frac32 - 3\epsilon} \,.
     \end{split}
\end{equation}
Each integral may be evaluated as a series in $\epsilon$, yielding,
\begin{equation}
    \begin{split}
    J\left(0,2,1,1,\tfrac12 + \epsilon\right)
    &= {\Gamma\left(\frac32 + 3\epsilon\right) (-p^2)^{-\frac32 - 3\epsilon} \over \Gamma\left(\frac12 + \epsilon\right) (4\pi)^{\D}}{-1\over 2\epsilon}\bigg\{
    {-1\over 3\epsilon} + \frac43 \log{2} + 2 
    + \left( {5\pi^2\over 9} - \frac83\log^2{2} - 8\log{2} - 12\right)\epsilon
    \\
    &\quad
    + \left[ -{62\zeta(3)\over 3} -{10\pi^2\over 3}  + 72 + {32\over 9}\log^3{2} + 16 \log^2{2} + \left( -{20\pi^2\over 9} + 48\right) \log{2} 
    \right]\epsilon^2 + \Order(\epsilon^3) \bigg\}\,.
    \end{split}
\end{equation}
%

\section{Wavefunction solution: Schrodinger-Coulomb \label{app:wf-schr}}
Consider the Lippmann-Schwinger equation and its related Born series for the solution of the Schrodinger equation
\begin{equation}\label{eq:LS}
    \psi_{\vb{p}}^{(\pm)}(\vb{x})= 
\langle \vb{x} | \left( 1 + \frac{1}{E-\hat{H}_0 \pm\iu0 } \hat{V} 
+ \frac{1}{E-\hat{H}_0\pm\iu0} \hat{V} \frac{1}{E-\hat{H}_0\pm\iu0} \hat{V} + \dots \right)
| \vb{p}\rangle \,,
\end{equation}
where $\hat{H}_0 = \hat{\vb{p}}^2/(2m)$ is the free Hamiltonian and $\hat{V}= V(\hat{\vb{x}})$ is the potential.   
For a finite range potential, the $+\iu 0$ ($-\iu 0$) prescription in \cref{eq:LS} corresponds to a plane wave plus  outgoing (incoming) spherical wave at large distance.  
Inserting a complete set of momentum eigenstates we arrive at
\begin{equation}
    \begin{split}
    \psi_{\vb{p}}^{(\pm)}(\vb{x})= \e^{\iu \vb{p} \cdot \vb{x}} \bigg[ 1 &+\int \frac{\dd^3 L}{(2\pi)^3} \e^{\iu \vb{L} \cdot \vb{x}} \frac{-2m}{2\vb{p}\cdot \vb{L} + \vb{L}^2\mp \iu 0} \tilde{V}(\vb{L}) \\
    &+  \int \frac{\dd^3 L_1}{(2\pi)^3} \frac{\dd^3 L_2}{(2\pi)^3} \e^{\iu \vb{L}_2 \cdot \vb{x}} \frac{-2m}{2\vb{p}\cdot \vb{L}_2 + \vb{L}_2^2\mp \iu 0} \tilde{V}(\vb{L}_2-\vb{L}_1) \frac{-2m}{2\vb{p}\cdot \vb{L}_1 + \vb{L}_1^2 \mp \iu 0} \tilde{V}(\vb{L}_1) + \dots  \bigg]~,
    \end{split}
\end{equation}
where $\tilde{V}(\vb{L}) = \int \dd^3x ~\e^{\iu \vb{L} \cdot\vb{x}} V(\vb{x})$ is the potential in momentum space. 
In particular, for a Yukawa potential,
$V(\vb{x})= (-Ze^2) \exp(-\lambda |\vb{x}|)/ (4\pi |\vb{x}|)$, we have
$\tilde{V}(\vb{L}) = -Ze^2/(\vb{L}^2 + \lambda^2)$.
Setting $\vb{x}\rightarrow 0$ and choosing the outgoing $+\iu 0$ prescription, the wavefunction $\psi_{\vb{p}}^{(+)}(0)$ provides an all-orders solution for the amplitude \cref{eq:SCgeneral}.   

Let us solve the Schrodinger equation, 
\begin{equation}
    \qty[-\frac{1}{2m} \nabla^2 - \frac{Z\alpha}{r}\e^{-\lambda r} ]\psi(\vb{x}) = \frac{\vb{p}^2}{2m} \psi(\vb{x})~,
\end{equation}
in the limit where $\lambda \ll |\vb{p}|$ (but to all orders in $Z\alpha$). 
Here $r=|\vb{x}|$. 
Let us write $\psi_{\vb{p}}(\vb{x}; \lambda)= \e^{\iu \vb{p} \cdot \vb{x}} F_{\vb{p}}(\vb{x},\lambda)$.  Choosing $\vb{p}$ along the $\hat{\vb{z}}$ direction, 
$\vb{p} = p \hat{\vb{z}}$, we look for the solution that reduces to 
$F=1$
at $z\to -\infty$ to obtain $\psi^{(+)}$,   
and the solution that reduces to $F=1$ at $z\to +\infty$ for $\psi^{(-)}$. 
The differential equation for $F$ is
\begin{equation}
    \label{matched-asym-1}
    \qty[-\frac{1}{2} \nabla^2 - \iu \vb{p}\cdot \nabla -   \frac{m Z\alpha}{r}\e^{-\lambda r} ]F(\vb{x}) =0 \,.
\end{equation}
We may now apply boundary layer theory~\cite{Bender1999}, solving for solutions at short and long distances and matching the solutions in their common domain of validity $p^{-1} \ll r \ll \lambda^{-1}$. For $r\ll \lambda^{-1}$, the Schrodinger equation is 
\begin{equation}
    \left[ -\frac12 {\nabla^2\over p^2} -\iu{\hat{\vb{p}} \cdot \vec{\nabla}\over p} - {\xi \over pr} \right]F_{<} = 0 \,,
\end{equation}
with solution 
\begin{align}
    F^{(+)}_{<}(\vb{x}) = N(p,\lambda) {}_1F_1(\iu\xi, 1, \iu p(r-z)) \,,
\end{align}
where $\!~_1F_1(a,b,c)$ is the confluent hypergeometric function. 
For $r\gg p^{-1}$, the Schrodinger equation is 
\begin{equation}
    \left[  -\iu{\hat{\vb{p}}\cdot \vec{\nabla}\over \lambda} - {\xi \over \lambda r}\e^{-\lambda r} \right]F_{>} = 0 \,,
\end{equation}
with solution 
\begin{align}
    F^{(+)}_{>}(\vb{x}) = \exp\bigg[\iu\xi \int_{-\infty}^z dz^\prime \,,
     {\e^{-\lambda \sqrt{z^{\prime 2} + r^2-z^2}} \over \sqrt{z^{\prime 2} + r^2-z^2} } \bigg]\,.
\end{align}
In the overlap region $p^{-1} \ll r \ll \lambda^{-1}$, the respective solutions can be expanded as 
\begin{align}
    F^{(+)}_{<} &\to N(p,\lambda) {1\over \Gamma(1-\iu \xi)} 
    \exp\bigg\{-{\pi \xi \over 2} -\iu\xi \log[p(r-z)] \bigg\} \,,
    \nl
    F^{(+)}_{>} &\to \exp\bigg\{\iu\xi \left[ -\log{\lambda (r-z) \over 2} - \gamma_{\rm E}\right] \bigg\} \,. 
\end{align}
Identifying $F^{(+)}_{<}=F^{(+)}_{>}$ in the overlap region, and using 
that ${}_1F_1(a,b,0) = 1$, we have, up to $\lambda/p$ power corrections,
\begin{align}
   \psi^{(+)}_{\vb{p}}(\vb{x}=0) = N(p,\lambda) &= \Gamma(1-\iu \xi) \exp\bigg\{ {\pi \over 2}\xi 
    + \iu \xi \left[ \log{2p\over\lambda} - \gamma_{\rm E}\right] \bigg\} \,.
\end{align}
The incoming solution $\psi^{(-)}_{\vb{p}}(\vb{x})$ is given by 
$F^{(-)}(\vb{x}) = [F^{(+)}(-\vb{x})]^*$.

\section{Wavefunction solution: Dirac-Coulomb \label{app:wf-dirac}}
The Dirac equation can be similarly shown to have a Lippmann-Schwinger solution and associated Born series. Let us define $\Phi(\vb{x})= u(\vb{p}) \e^{\iu \vb{p} \cdot \vb{x}}$, where $u(\vb{p})$ is a Dirac spinor.  
The solution of the Dirac equation with a potential can be written as
\begin{equation}
    \begin{split}
    \psi^{(\pm)}(\vb{x}) = \bigg[1  &+ \int \frac{\dd^3 L}{(2\pi)^3}  
    ~\e^{\iu \vb{L} \cdot \vb{x}} \frac{1}{\slashed{p}+ \slashed{L}-m\pm\iu 0 } \gamma_0  \tilde{V}(\vb{L})   \\
    &+ \int \frac{\dd^3 L_2}{(2\pi)^3}   \frac{\dd^3 L_1}{(2\pi)^3}   ~\e^{\iu \vb{L}_2 \cdot \vb{x}} \frac{1}{\slashed{p}+ \slashed{L}_2-m\pm\iu 0 } \gamma_0 \tilde{V}(\vb{L}_1-\vb{L}_2)  \frac{1}{\slashed{p}+ \slashed{L}_1-m\pm\iu 0 } \gamma_0 \tilde{V}(\vb{L}_1)   + ... \bigg]\Phi(\vb{x})\,.
    \end{split}
\end{equation}
The amplitude of interest, \cref{eq:DC-xreg}, is given by 
$\bar{u}(\vb{p}) {\cal  M}_{\vb{r}} = \overline{\psi}^{(-)}(-\vb{r}) = [\psi^{(-)}(-\vb{r})]^\dagger \gamma^0$.  
%
%
We require the solution $\psi^{(-)}$ with a small but non-zero photon mass $\lambda$. References~\cite{Rose:1937xxx,Jackson:1957auh} present the angular momentum components for the strict $\lambda=0$ solution, which is related to our problem by a normalization that must be computed.

To determine the complete solution including $\lambda$ dependence, we identify this solution with $\psi_<^{(-)}$, up to a normalization that is fixed by matching to $\psi_>^{(-)}$ in the overlapping region of validity.  For simplicity we perform the matching by projecting onto the $S$-wave component of the outgoing spherical wave.

Let us consider the upper components of $\psi^{(\pm)}$ in the Dirac basis for $\gamma^\mu$, and introduce
\begin{align}
    {1+\gamma_0\over 2} \psi^{(\pm)}(\vb{x}) &= \e^{\iu \vb{p}\cdot \vb{x}} 
    F_{\vb{p}}^{(\pm)}(\vb{x},\lambda) \left(\begin{array}{c} \chi \\ 0 \end{array}\right) \,,
\end{align}
where $\chi$ is a 2-component spinor.  Similar to the Schrodinger-Coulomb problem, we look for solutions $F_>$ when $r \gg p^{-1}$, and $F_<$ when $r \ll \lambda^{-1}$.

The large-distance solution obeys an identical equation to the Schrodinger-Coulomb problem (with $\xi=Z\alpha/\beta$ and $\beta =p/E$ representing the relativistic velocity).   The solution for $F_>^{(-)}$ is given in Appendix~\ref{app:wf-schr}, and for the matching we require the small-$r$ limit. Considering the outgoing spherical wave component, the $S$-wave projection is
\begin{align}\label{eq:Fbig_small}
    \psi_>^{(-)} \to {\e^{ipr}\over 2\iu pr} \exp\left[ -\iu \xi \left( \log{2p\over\lambda} - \gamma_E  \right) + \iu \xi \log(2pr) \right] \,.
\end{align}
The relevant component of the small-$r$ solution involves the quantity~\cite{Rose:1937xxx,Jackson:1958xxx}
\begin{align} 
    C(p,\lambda)\, f_{-1}(pr) &= C(p,\lambda)\, \e^{{\pi \xi \over 2}} {|\Gamma(\eta + \iu \xi)|\over \Gamma(2\eta + 1)} (2pr)^{\eta-1}
    \bigg\{ 
    \e^{-\iu pr + \iu \kappa}(\eta + \iu \xi) {}_1F_1(\eta+1+ \iu \xi, 2\eta+1, 2\iu pr) 
    + {\rm c.c.} 
    \bigg\} \,,
\end{align}
where $\exp(\iu\kappa)=\sqrt{(1+\iu m\xi/E)/(\eta+\iu \xi)}$.
From the large-$r$ limit of this expression, taking the outgoing spherical wave component, we have 
\begin{align}\label{eq:Fsmall_big}
    \psi_<^{(-)}&\to C(p,\lambda) {|\Gamma(\eta+\iu \xi)| \over \Gamma(\eta+\iu \xi)}
    {\e^{\iu pr}\over 2\iu p r} \exp\left[ \iu \xi \log(2pr) - \iu (\eta-1){\pi \over 2} + \iu \kappa  \right] \,.
\end{align}
Comparison of \cref{eq:Fbig_small,eq:Fsmall_big} in the overlap region $p^{-1} \ll r \ll \lambda^{-1}$ determines $C(p,\lambda)$.   
Using  ${}_1\!F_1(a,b,0) = 1$,
and taking the $r\to 0$ limit of the complete solution, we have ({\it cf.} Eq.~(16) of Ref.~\cite{Jackson:1958xxx})
\begin{equation}
    \begin{split}
    \label{wavefunction-Jackson-plus-phase}
   \lim_{r\rightarrow 0}\psi^{(-)}(\vb{x}; \lambda) = \e^{\iu \xi[-\log(2p/\lambda)+\gamma_{\rm E}]}
   \e^{\pi \xi/2}\Gamma(\gammatrad +\iu \xi) \times \frac{1+\gammatrad +\iu \xi \qty(1-\tfrac{m}{E}) }{\Gamma(1+2\gammatrad)}\e^{-\iu(1-\gammatrad)\pi/2}
   (2pr)^{\eta-1} 
   \\
  \times \bigg[ 1 + \frac{ Z\alpha}{1+\gammatrad } \frac{\iu \gamma_0 \vb*{\gamma}\cdot\vb{x}}{|\vb{x}|}\bigg] \bigg[\qty(\frac{1+M}{2}) + \qty(\frac{1-M}{2})\gamma_0 \bigg]  u(\vb{p})\,,
  \end{split}
\end{equation}
where 
\begin{align}
      M&= \frac{E+m}{E+\gammatrad m} \qty( 1+ \iu \xi \frac{m}{E} ) \,.
\end{align}
%

\section{All orders UV function with a finite-distance regulator \label{app:all-orders-xreg} } 
We can compute the UV matching coefficient introduced in \cref{eq:M-r} by setting $\lambda=p=0$ 
and evaluating the remaining integrals using dimensional regularization. Examining the perturbative series we find that the (bare, unrenormalized) 
UV matrix element has the following structure 
\begin{equation}\label{eq:Fidef}
     \mathcal{M}_{\rm{UV}}^{\rm bare} = F_1^{\rm bare} - F_2^{\rm bare} 
     \times \frac{\iu \gamma_0\vb*{\gamma}\cdot\vb{x}}{2|\vb{x}|} ~,
\end{equation}
where 
\begin{equation} 
    F_1^{\rm bare} = 
    \sum_{n=0}^\infty (Z e^2)^{2n}  \mathcal{I}_1^{(n)} \,, \quad 
    F_2^{\rm bare} \times \frac{\iu \gamma_0\vb*{\gamma}\cdot\vb{x}}{2|\vb{x}|} = (-1)\times \sum_{n=0}^\infty (Z e^2)^{2n+1} \mathcal{I}_2^{(n)} \,.
\end{equation}
In particular, all even orders of perturbation theory contribute to $F_1$ and all odd orders to $F_2$. 
The lowest order loop integrals are given by
$\mathcal{I}_1^{(0)}=1$, 
\begin{align}
    \mathcal{I}_1^{(1)}&=\int \frac{\dd^\D L_{1}}{(2\pi)^\D} \frac{\dd^\D L_{2}}{(2\pi)^\D} ~\e^{-\iu \vb{L}_{2} \cdot\vb{x} }~\frac{\gamma_0\vb*{\gamma}\cdot \vb{L}_2}{\vb{L}_2^2}\frac{1}{(\vb{L}_2-\vb{L}_1)^2}\frac{\gamma_0\vb*{\gamma}\cdot \vb{L}_1}{\vb{L}_1^2} \frac{1}{\vb{L}_1^2} \,, 
    \nl
    \mathcal{I}_2^{(0)}&=\int \frac{\dd^\D L_{1}}{(2\pi)^\D}  ~\e^{-\iu \vb{L}_{1} \cdot\vb{x} } ~\frac{\gamma_0\vb*{\gamma}\cdot \vb{L}_1}{\vb{L}_1^2} \frac{1}{\vb{L}_1^2} \,,
\end{align}
and for higher orders,
\begin{align}
    \mathcal{I}_1^{(n)}&=\int \frac{\dd^\D L_{2n}}{(2\pi)^\D}~ \e^{-\iu \vb{L}_{2n} \cdot\vb{x} } \frac{\gamma_0\vb*{\gamma}\cdot \vb{L}_{2n}}{\vb{L}_{2n}^2}\qty[\prod_{i=2}^{2n-1} \int\frac{\dd^\D L_{i}}{(2\pi)^\D}
    \frac{\gamma_0\vb*{\gamma}\cdot \vb{L}_i}{\vb{L}_i^2} \frac{1}{(\vb{L}_i-\vb{L}_{i+1})^2}] 
    \int \frac{\dd^\D L_{1}}{(2\pi)^\D}  \frac{1}{(\vb{L}_2-\vb{L}_1)^2}\frac{\gamma_0\vb*{\gamma}\cdot \vb{L}_1}{\vb{L}_1^2} \frac{1}{\vb{L}_1^2}~, \\
    \mathcal{I}_2^{(n)}&=\int \frac{\dd^\D L_{2n+1}}{(2\pi)^\D}~ \e^{-\iu \vb{L}_{2n+1} \cdot\vb{x} } \frac{\gamma_0\vb*{\gamma}\cdot \vb{L}_{2n+1}}{\vb{L}_{2n+1}^2}\qty[\prod_{i=2}^{2n} \int\frac{\dd^\D L_{i}}{(2\pi)^\D}
    \frac{\gamma_0\vb*{\gamma}\cdot \vb{L}_i}{\vb{L}_i^2} \frac{1}{(\vb{L}_i-\vb{L}_{i+1})^2}] 
    \int \frac{\dd^\D L_{1}}{(2\pi)^\D}  \frac{1}{(\vb{L}_2-\vb{L}_1)^2}\frac{\gamma_0\vb*{\gamma}\cdot \vb{L}_1}{\vb{L}_1^2} \frac{1}{\vb{L}_1^2}\,. \nonumber
\end{align}
These integrals are recursively one-loop, and can be evaluated by repeated use of the following identity:
\begin{equation}
    \begin{split}
    C(\nu_j) 
    \frac1{(\vb{L}_{2j+1}^2)^{\nu_{j+1}-1}}
    &= \int \frac{\dd^\D L_{2j-1}}{(2\pi)^\D}\frac{\dd^\D L_{2j}}{(2\pi)^\D}   \frac{1}{(\vb{L}_{2j+1}-\vb{L}_{2j})^2} 
    \frac{\gamma_0\vb*{\gamma}\cdot \vb{L}_{2j}}{\vb{L}_{2j}^2} 
    \frac{1}{(\vb{L}_{2j}-\vb{L}_{2j-1})^2} \frac{\gamma_0\vb*{\gamma}\cdot \vb{L}_{2j-1}}{\qty(\vb{L}_{2j-1}^2)^{\nu_j}}\\
    &= \frac{1}{(4\pi)^\D} \frac{\Gamma(\nu_j+2-\D)}{\Gamma(\nu_j)} \betaFunc(\tfrac{\D}{2}-1,1+\tfrac{\D}{2}-\nu_j)
    \betaFunc(\tfrac{\D}{2}-1,\D-\nu_j-1)\qty(\frac1{\vb{L}_{2j+1}^2})^{\nu_j+2-\D} \,,
    \end{split} 
\end{equation}
where $\nu_1=2$ and $\nu_{j+1}=\nu_j+3-\D$ so that $\nu_j= 2 + 2(j-1)\epsilon$. The final integral involving $\e^{-\iu \vb{L}_{2n} \cdot \vb{x}}$ for $\mathcal{I}_1^{(n)}$  
(or $\e^{-\iu \vb{L}_{2n+1} \cdot \vb{x}}$ for $\mathcal{I}_2^{(n)}$)
can be evaluated with a Schwinger parameter, yielding 
\begin{equation}
    \begin{split}
    \mathcal{I}_1^{(n)}&=\qty[\prod_{j=1}^{n-1} C(\nu_j)] \times \frac{\Gamma(\D-\nu_n-1)}{(4\pi)^\D\Gamma(\nu_n)}\betaFunc(\tfrac{\D}{2}-1,1+\tfrac{\D}{2}-\nu_n)  \qty(\frac{\vb{x}^2}{4})^{\nu_n+1-\D}~, \\
    \mathcal{I}_2^{(n)}&= \Bigg[\prod_{j=1}^{n} C(\nu_j)\Bigg]
   \qty[ \frac{2\Gamma(\tfrac{\D}{2} -\nu_{n+1}+1)}{(4\pi)^{\D/2}\Gamma(\nu_{n+1}) }]\qty[\frac{\vb{x}^2}{4}]^{\nu_{n+1}-(\D+1)/2}\times ~ \frac{-\iu \gamma_0 \vb*{\gamma}\cdot \vb{x}}{2|\vb{x}|}\,.
    \end{split}
\end{equation}
Using the properties of the Gamma function, 
the functions $F_1$ and $F_2$ can be shown to have the following series expansion
\begin{equation}
    \begin{split}
    F_1^{\rm bare}&= \sum_{n=0}^\infty \tilde{g}^n 
    \frac{(-1)^n}{n!} \qty(\frac{1}{\epsilon})^n  ~\qty[\prod_{m=0}^{n-1} \frac{1}{(1+2 m \epsilon)}]
    \,, \\
    F_2^{\rm bare}&= Z\widetilde{\alpha} \sum_{n=0}^\infty \tilde{g}^n 
    \frac{(-1)^n}{n!} \qty(\frac{1}{\epsilon})^n  ~\qty[\prod_{m=0}^{n} \frac{1}{(1+2 m \epsilon)}]
    \label{series2}  \,,
    \end{split}
\end{equation}
where in terms of $\overline{\alpha}(\mu)$ in \cref{eq:msbar} we define
\begin{align}
    \tilde{g} = {(Z\widetilde{\alpha})^2\over 8} 
\end{align}
with
\begin{align}
    \widetilde{\alpha} = \overline{\alpha}\left(\mu^2 r^2\over 16\right)^\epsilon {\Gamma(\frac12-\epsilon)\over \Gamma(\frac12+\epsilon)}
     = \overline{\alpha}\left(\mu r \e^{\gamma_{\rm E}} \right)^{2\epsilon} \left[ 1 + {\cal O}(\epsilon^2) \right]
   \,. 
\end{align}
In particular, when expressed in terms of $\widetilde{\alpha}$, the coefficients in the 
perturbative expansion of $F_i^{\rm bare}$
are expressible entirely as rational functions of $\epsilon$. Choosing $\mu = (r\e^{\gamma_{\rm E}})^{-1}$ so that $\widetilde{\alpha}$ can be identified with the $\overline{\rm MS}$ coupling, we find that the $\overline{\rm MS}$ operator renormalization constant can be written as $\exp[ \frac{1}{\epsilon}\sum_n a_n \tilde{g}^n]$ for some numbers\footnote{
The leading orders 
obtained by direct evaluation from \cref{series2}
are 
\begin{equation}
    \mathcal{Z} = \exp\qty[ \frac1\epsilon \qty(\tilde{g}+ \tilde{g}^2+\frac{8 \tilde{g}^3}{3}+10 \tilde{g}^4+\frac{224 \tilde{g}^5}{5}+224 \tilde{g}^6+\frac{8448 \tilde{g}^7}{7}+6864 \tilde{g}^8+\frac{366080 \tilde{g}^9}{9}+ \frac{1244672 \tilde{g}^{10}}{5}~) + \ldots ] \,.
\end{equation}
We have checked explicitly to 16$^{\rm th}$ order in $\tilde{g}$ that the renormalization constant can be written as $\exp[ \frac{1}{\epsilon}\sum_n a_n \tilde{g}^n]$, consistent with the explicit all-orders expressions in \cref{F2-asym,F1-asym}.
\label{Catalan}}
$a_n$. The sequence of coefficients can be related to the Catalan numbers $\mathcal{C}(n)=(2n)!/(n!(n+1)!)$.\!\footnote{We were able to identify this sequence with help from the Online Encyclopedia of Integer Sequences \cite{OEIS:ref}. \label{OEIS}} The series in the exponent then converges, and is given by 
\begin{equation}
    \log(\mathcal{Z}) = \frac{1}{\epsilon}\sum_{n=0}^\infty \frac{2^n\mathcal{C}(n)}{n+1} \tilde{g}^{n+1} = \frac{1}{2\epsilon} \left[-\sqrt{1-8 \tilde{g}}+\log \left(\sqrt{1-8 \tilde{g}}+1\right)+1-\log (2)\right] \,.
    \label{log-Z}
\end{equation}
The series in \cref{series2} can also be summed, and converges for any nonzero $\epsilon$. The answer is given by
\begin{align}
     F_1^{\rm bare} &=  2^{\frac{1}{4 \epsilon }-\frac{1}{2}} \left(\frac{\sqrt{\tilde{g}}}{\epsilon }\right)^{1-\frac{1}{2 \epsilon }} \Gamma \left(\frac{1}{2 \epsilon }\right) J_{\frac{1}{2 \epsilon }-1}\left(\frac{\sqrt{8} \sqrt{\tilde{g}}}{\epsilon }\right)~,\label{F1-all-orders}\\
    (Z\widetilde{\alpha})^{-1} F_2^{\rm bare} &= 2^{\frac{1}{4 \epsilon }} \left(\frac{\sqrt{\tilde{g}}}{\epsilon }\right)^{-\frac{1}{2 \epsilon }} \Gamma \left(1+\frac{1}{2 \epsilon }\right) J_{\frac{1}{2 \epsilon }}\left(\frac{\sqrt{8} \sqrt{\tilde{g}}}{\epsilon }\right)~ \label{F2-all-orders}.
\end{align}
Using \cref{F1-all-orders,F2-all-orders} we can see how renormalization works at all orders in the  coupling. We require the $\epsilon\rightarrow 0$ asymptotic behavior of the Bessel functions. The relevant identity is ({\it cf.} Eq.\ (10.20.4) of Ref.~\cite{NIST:DLMF})
\begin{equation}
    \lim_{\nu\rightarrow \infty} J_\nu(\nu z) \sim \frac{\sqrt[4]{\frac{4 \zeta (z)}{1-z^2}} \text{Ai}\left(\nu ^{2/3} \zeta (z)\right)}{\sqrt[3]{\nu }} \qq{with} \zeta(z) = \qty[\frac{3}{2} \left(-\sqrt{1-z^2}+\log \left(\sqrt{1-z^2}+1\right)-\log (z)\right)]^{2/3}~,
\end{equation}
where we adopt the notation of Ref.~\cite{NIST:DLMF} and use $\sim$ to denote ``asymptotic to''. Using this identity, Sterling's approximation, and the large argument limit of the Airy function it is straightforward to show that 
\begin{equation}
    \label{F2-asym}
     (Z\widetilde{\alpha})^{-1}  F_2^{\rm bare} \sim  \qty(\frac{1}{1-8 \tilde{g}})^{1/4} ~\exp\qty[\frac{1}{2\epsilon} \left(\sqrt{1-8 \tilde{g}}-\log \left(\sqrt{1-8 \tilde{g}}+1\right)-1+\log (2)\right)]~\qq{as} \epsilon \rightarrow 0\,.
\end{equation}
For $F_1$ it is convenient to introduce $1/2\epsilon' = 1- 1/2\epsilon$ and $\tilde{g}' = \tilde{g}(1+2\epsilon')$.  We then find 
\begin{equation}
    \label{F1-asym}
     F_1^{\rm bare}\sim  \qty(\frac{1}{1-8 \tilde{g}'})^{1/4} ~\exp\qty[\frac{1}{2\epsilon'}\left(\sqrt{1-8 \tilde{g}'}-\log \left(\sqrt{1-8 \tilde{g}'}+1\right)-1+\log (2)\right)]~\qq{as} \epsilon \rightarrow 0\,.
\end{equation}
Notice that the form of $\mathcal{Z}$ that we obtained from recognizing the infinite sequence using our perturbative 
result is precisely what is needed for all-orders renormalization, {\it cf.} \cref{log-Z,F2-asym}. We find 
\begin{align}\label{eq:mu0}
    F_1|_{\mu=(r\e^{\gamma_{\rm E}})^{-1}}&= \lim_{\epsilon \rightarrow 0} \mathcal{Z} F_1^{\rm bare} = \frac{\sqrt{1-(Z\alpha)^2} +1}{2} 
    \qty(\frac{1}{1-(Z\alpha)^2})^{1/4}~,\\
    F_2|_{\mu=(r\e^{\gamma_{\rm E}})^{-1}}&= \lim_{\epsilon \rightarrow 0} \mathcal{Z} F_2^{\rm bare} = Z\alpha \qty(\frac{1}{1-(Z\alpha)^2})^{1/4}\,.
\end{align}
The $\mu$ dependence of the renormalized coefficient functions $F_i$ is governed by the anomalous dimension, 
\begin{align}\label{eq:dlogmu}
    {\dd \over \dd \log\mu} F_i &= \gamma_\mathcal{O} F_i \,, 
\end{align}
and $\gamma_\mathcal{O}$ is determined by the coefficient of $1/\epsilon$ in the corresponding $\overline{\rm MS}$ operator renormalization constant: 
\begin{align}
\mathcal{Z} = \sum_{m=0}^{\infty} {1\over \epsilon^m} \mathcal{Z}_m \,, \quad 
 \gamma = - 2\alpha {\partial \over \partial \alpha} \mathcal{Z}_1   \,.
\end{align}
Using the explicit form of $\mathcal{Z}$ we have, to all orders in the coupling,
\begin{align}
 \mathcal{Z}_1 = \frac12 \bigg[ 1 - \sqrt{1-(Z\alpha)^2} + \log{1 + \sqrt{1-(Z\alpha)^2}\over 2}\bigg] \,, 
\end{align}
and so taking the derivative, {\it cf.} \cref{anom-dim}, we find
\begin{align}
 \gamma_{\mathcal{O}}= \sqrt{1-(Z\alpha)^2} -1 \,.
\end{align}
Using the solution of \cref{eq:dlogmu} with 
initial condition \cref{eq:mu0}, 
the amplitude (\ref{eq:Fidef}) after $\overline{\rm MS}$ renormalization is 
\begin{equation}
    \label{MH-final}
    \begin{split}
    \mathcal{M}_{\rm UV}^R(\mu) &= \qty(\mu r \e^{\gamma_{\rm E}})^{\gammatrad -1}\frac{1+\gammatrad }{2\sqrt{\gammatrad }} \qty[ 1 - \frac{Z\alpha}{1+\gammatrad } \frac{\iu \gamma_0 \vb*{\gamma}\cdot \vb{x}}{|\vb{x}|}]~,
    \end{split}
\end{equation}
where we have used $\gammatrad=\sqrt{1-(Z\alpha)^2}$.

\newpage

\bibliography{fermi.bib}

\end{document}